\documentclass{SciPost}

\hypersetup{
    colorlinks,
    linkcolor={red!50!black},
    citecolor={blue!50!black},
    urlcolor={blue!80!black}
}
\usepackage[normalem]{ulem}
\usepackage[bitstream-charter]{mathdesign}
\usepackage[capitalize]{cleveref}

\DeclareSymbolFont{usualmathcal}{OMS}{cmsy}{m}{n}
\DeclareSymbolFontAlphabet{\mathcal}{usualmathcal}

\fancypagestyle{SPstyle}{
\fancyhf{}
\lhead{\colorbox{scipostblue}{\bf \color{white} ~SciPost Physics Lecture Notes }}
\rhead{{\bf \color{scipostdeepblue} ~Submission }}

\fancyfoot[C]{\textbf{\thepage}}
}

\newcommand{\s}{\newline \vspace*{-3.5mm}}
\newcommand{\beq}{\begin{eqnarray}}
\newcommand{\eeq}{\end{eqnarray}}
\newcommand{\mathcalM}{\mathcal{M}}
\newcommand{\mathcalR}{\mathcal{R}}

\newcommand{\diag}{\text{diag}}

\renewcommand{\Re}{\text{Re}\,}

\usepackage{xspace}
\usepackage{units}

\newcommand{\BSMArt}{{\tt BSMArt}\xspace}
\newcommand{\NMSSMCALC}{{\tt NMSSMCALC}\xspace}

\begin{document}

\begin{flushright}
\texttt{DESY-26-056}\\
\texttt{FR-PHENO-2026-008}\\
\texttt{KA-TP-09-2026}
\end{flushright}

\pagestyle{SPstyle}

\begin{center}{\Large \textbf{\color{scipostdeepblue}{
\texttt{NMSSMScanner}: Efficient Scans in the NMSSM Parameter Space \\[0.1cm] Proof of Concept
}}}\end{center}

\begin{center}\textbf{
Rafael Boto\textsuperscript{1$\star$},
Thi Nhung Dao\textsuperscript{2$\dagger$},
Felix Egle\textsuperscript{3$\ddagger$},
Karim Elyaouti\textsuperscript{1$\S$},
Martin Gabelmann\textsuperscript{4$\P$},
Margarete M\"uhlleitner\textsuperscript{1$\parallel$},
Johann Plotnikov\textsuperscript{1$\star\star$}
}\end{center}

\begin{center}
{\bf 1} Institute for Theoretical Physics, Karlsruhe Institute of Technology,
Wolfgang-Gaede-Str. 1, 76131 Karlsruhe, Germany
\\  
{\bf 2} Phenikaa Institute for Advanced Study, PHENIKAA University,
Hanoi 12116, Vietnam
\\
{\bf 3} Deutsches Elektronen-Synchrotron DESY,
Notkestr.~85, 22607 Hamburg, Germany
\\
{\bf 4} Albert-Ludwigs-Universität Freiburg, Physikalisches Institut,
Hermann-Herder-Str. 3, 79104 Freiburg, Germany
\\[\baselineskip]
$\star$ \href{mailto:rafael.boto@kit.edu}{\small rafael.boto@kit.edu}\,,
$\dagger$ \href{mailto:nhung.daothi@phenikaa-uni.edu.vn}{\small nhung.daothi@phenikaa-uni.edu.vn}\,,
$\ddagger$ \href{mailto:felix.egle@desy.de}{\small felix.egle@desy.de}\,, 
\\ $\S$ \href{mailto:karim.elyaouti@partner.kit.edu}{\small karim.elyaouti@partner.kit.edu}\,,
$\P$ \href{mailto:martin.gabelmann@physik.uni-freiburg.de}{\small
  martin.gabelmann@physik.uni-freiburg.de}\,, \\
$\parallel$ \href{mailto:margarete.muehlleitner@kit.edu}{\small margarete.muehlleitner@kit.edu}\,,
$\star\star$ \href{mailto:johann.plotnikov@partner.kit.edu}{\small johann.plotnikov@partner.kit.edu}
\end{center}

\section*{\color{scipostdeepblue}{Abstract}}
We present the first version of the new scanning tool \texttt{NMSSMScanner} that allows to perform efficient scans in the complex multi-parameter space of the Next-to-Minimal Supersymmetric extension of the Standard Model (NMSSM) while taking into account all relevant constraints. As a proof of concept we apply it to the search for NMSSM parameter configurations that maximize Higgs boson pair production from resonant scalar or pseudoscalar production in various final states.

\vspace{\baselineskip}

\noindent\textcolor{white!90!black}{%
\fbox{\parbox{0.975\linewidth}{%
\textcolor{white!40!black}{\begin{tabular}{lr}%
  \begin{minipage}{0.6\textwidth}%
    {\small Copyright attribution to authors. \newline
    This work is a submission to SciPost Physics Lecture Notes. \newline
    License information to appear upon publication. \newline
    Publication information to appear upon publication.}
  \end{minipage} & \begin{minipage}{0.4\textwidth}
    {\small Received Date \newline Accepted Date \newline Published Date}%
  \end{minipage}
\end{tabular}}
}}
}

\nolinenumbers

\vspace{10pt}
\noindent\rule{\textwidth}{1pt}
\tableofcontents
\noindent\rule{\textwidth}{1pt}
\vspace{10pt}


\section{Introduction}
\label{sec:intro}
Open questions like the nature of Dark Matter (DM) or why there is more matter than antimatter in the universe call for extensions of the Standard Model (SM) of particle physics. Among these, supersymmetry (SUSY) is particularly compelling, as it not only solves (some of) the open problems but also relates bosons and fermions through SUSY transformations. Supersymmetry requires the introduction of at least two complex Higgs doublets, as realized in the Minimal Supersymmetric extension of the SM (MSSM). The next-to-MSSM (NMSSM) furthermore adds a complex singlet superfield and thereby solves the so-called $\mu$ problem of the MSSM, at the price of an enlarged set of input parameters. Supersymmetry implies an upper bound on the tree-level mass of the lightest doublet-like CP-even Higgs boson, so that higher-order corrections have to be included in order to comply with the measured \unit[125]{GeV} mass of the discovered Higgs boson. Consequently, the Higgs boson masses are not input parameters anymore, but derived quantities. Together with the fact that the model depends on a large number of input parameters, this makes the scans in the NMSSM parameter space notoriously difficult.
 The added complexity is particularly problematic when searching for
 benchmark scenarios that offer desired features such as specific mass
 and coupling configurations and at the same time fulfill all relevant
 theoretical and experimental constraints. \s
 
 In this paper, we present a
 new framework that allows for efficient scans of the NMSSM parameter
 space while taking into account all relevant collider and low-energy
 observables as well as Dark Matter constraints. As a
 proof of concept, we derive viable benchmark scenarios for the 
   resonant production of a SM--like plus non-SM-like Higgs boson pairs
   in various final states. 
The framework is based on a set of codes, that calculate the relevant
NMSSM observables and check for their compatibility with theoretical
and experimental constraints. For the parameter scan, different
strategies such as random scans, Markov-Chain-Monte-Carlo (MCMC), machine learning, or other custom algorithms can be employed with the help of \BSMArt \cite{Goodsell:2023iac}. The setup thereby allows for convenient scans of the NMSSM parameter space and efficiently finds viable and phenomenologically relevant benchmark scenarios as a necessary input for far- and near-future new physics searches of all kinds. \s

The remainder of this paper is organized as follows. In Sec.~\ref{sec:model} we briefly introduce the NMSSM to set our notation. In Sec.~\ref{sec:code}, we specify which observables are tested.  Section \ref{sec:results} applies the new scan framework focusing on the maximization of Higgs boson pair production cross sections, thereby providing selected benchmark points. We summarize in Sec.~\ref{sec:summary}.

\section{The NMSSM \label{sec:model}}
We work in the framework of the CP-violating NMSSM with a scale-invariant superpotential applying a discrete $\mathbb{Z}_3$ symmetry. We focus here on the presentation of the Higgs sector of the model. Further details and discussions of the complete NMSSM Lagrangian can be found e.g.~in \cite{Maniatis:2009re,Ellwanger:2009dp}. The Higgs potential is given by the sum of the $F$-term of the superpotential, the soft SUSY breaking Lagrangian and the $D$-term contributions. The scale-invariant NMSSM superpotential, which is added to
the MSSM superpotential $W^{\text{MSSM}}$, reads 
\begin{align}
W^{\text{NMSSM}}&=- \epsilon_{ij} \lambda \hat{S} \hat{H}^i_d \hat{H}^j_u + \frac{\kappa}{3}
\hat{S}^3+ W^{\text{MSSM}}\,, \quad \mbox{with} \nonumber\\
W^{\text{MSSM}}&=\epsilon_{ij} [ -  y_u \widehat{H}_u^i
\widehat{Q}^j \widehat{U}^c + y_d \widehat{H}_d^i \widehat{Q}^j
\widehat{D}^c  + y_e \widehat{H}_d^i \widehat{L}^j 
\widehat{E}^c ]\; ,
\label{eq:nmssmsuperpot}
\end{align}
where $\hat{H}_d$ and $\hat{H}_u$ denote the two Higgs doublet
superfields, $\hat{S}$ the singlet superfield, $\widehat{Q}$ and
$\widehat{L}$ the quark and leptonic left-handed doublet superfields,
respectively, and $\widehat{U}$, $\widehat{D}$, and $\widehat{E}$ the
corresponding right-handed singlet quark and lepton
superfields. They are understood to represent all
  three fermion families.
 The superscript $c$ denotes charge conjugation and $\epsilon_{ij}$ ($i,j=1,2$) is the totally antisymmetric tensor with $\epsilon_{12}= \epsilon^{12}= 1$ and $i,j$ denoting the indices of the fundamental $SU(2)_L$ representation. Here and in the following, we sum over repeated indices and suppress, for simplicity, color and generation indices. We neglect flavor mixing and assume the Yukawa couplings $y_u$, $y_d$ and $y_e$ to be diagonal $3\times 3$ matrices in flavor space. We can then reabsorb complex phases to render all SM fermion masses real by redefining the quark fields, without affecting physical observables. The dimensionless NMSSM-specific couplings $\lambda$
and $\kappa$ are in general complex in the CP-violating NMSSM. The cubic term in $\hat{S}$ breaks the Peccei-Quinn symmetry, avoiding a massless axion. The soft SUSY breaking Lagrangian reads
\begin{align}
\label{eq:Lagmass}
 {\cal L}_{\text{soft,NMSSM}} =&
 - m_{H_u}^2 | H_u |^2 -  m_{H_d}^2 | H_d|^2 - m_{{\widetilde
     Q}_3}^2|{\widetilde Q}_3^2|-  m_{\widetilde t_R}^2 |{\widetilde t}_R^2|
 -  m_{\widetilde b_R}^2|{\widetilde b}_R^2| -
 m_{{\widetilde L}_3}^2|{\widetilde L}_3^2| \nonumber \\
& - m_{\widetilde  \tau_R}^2|{\widetilde \tau}_R^2|
+ \epsilon_{ij} (y_t A_t H_u^i \widetilde Q_3^j \widetilde t_R^c - y_b A_b H_d^i 
\widetilde Q_3^j  \widetilde  b_R^c - y_\tau A_\tau H_d^i \widetilde L_3^j \widetilde \tau_R^c + \mathrm{h.c.})
\nonumber \\
& - \frac{1}{2} \bigg( M_1 \widetilde{B}
\widetilde{B} + M_2 \sum_{i=1}^3 \widetilde{W}_i \widetilde{W}_i +
M_3 \sum_{a=1}^8 \widetilde{G}_a \widetilde{G}_a  \ + \ {\rm h.c.}
\bigg) \nonumber \\
& - m_S^2 |S|^2 + (\epsilon_{ij}\lambda A_\lambda S H_d^i H_u^j - \frac{1}{3}
\kappa  A_\kappa S^3 + \mathrm{h.c.}) \;,
\end{align}
where for simplicity only the third generation of (s)fermions is displayed. The tilde over the fields denotes the superpartner of the respective SM field. The soft SUSY
breaking gaugino parameters $M_k$ ($k=1,2,3$) of the bino, wino and
gluino fields $\widetilde{B},$ $\widetilde{W}$ and $\widetilde{G}$, as
well as the soft SUSY breaking trilinear couplings $A_x$ ($x=\lambda,
\kappa, t, b, \tau$) are complex, whereas the soft SUSY
breaking mass parameters of the scalar fields, $m_X^2$
($X=S,H_d,H_u,\widetilde{Q}_3, \widetilde{t}_R, \widetilde{b}_R,
\widetilde{L}_3, \widetilde{\tau}_R$) are real. Applying the $R$-symmetry transformation, either $M_1$ or $M_2$ can be chosen real. We keep them both complex in the CP-violating NMSSM.
In the following $\lambda$ and $\tan\beta$
will be chosen positive by convention, whereas $\kappa, A_\lambda$ and $A_\kappa$ can take both signs. The final Higgs potential at tree level reads
\begin{align}
    V_H &= (|\lambda S|^2 + m_{H_d}^2)H_d^\dagger H_d+ (|\lambda S|^2
+ m_{H_u}^2)H_u^\dagger H_u +m_S^2 |S|^2 \nonumber \\
& + \frac{1}{8} (g_2^2+g_1^{2})(H_d^\dagger H_d-H_u^\dagger H_u )^2
+\frac{1}{2} g_2^2|H_d^\dagger H_u|^2 \label{eq:higgspotential} \\ 
&   + |-\epsilon^{ij} \lambda  H_{d,i}  H_{u,j} + \kappa S^2 |^2+
\big[-\epsilon^{ij}\lambda A_\lambda S   H_{d,i}  H_{u,j}  +\frac{1}{3} \kappa
A_{\kappa} S^3+\mathrm{h.c.} \big] \;,
\nonumber
\end{align}
where $g_1$ and $g_2$ denote the $U(1)_Y$ and $SU(2)_L$ gauge
couplings, respectively. After electroweak symmetry breaking (EWSB), the Higgs fields are expanded around their vacuum expectation values (VEVs) $v_u$,
$v_d$, and $v_S$, respectively, where two more CP-violating phases, $\varphi_u$ and $\varphi_s$, are introduced such that 
\begin{equation}
    H_d =
 \left( \begin{array}{c} \frac{1}{\sqrt 2}(v_d + h_d +i a_d)\\ h_d^- \end{array} \right) ,\;
H_u = e^{i\varphi_u}\left( \begin{array}{c}
h_u^+ \\ \frac{1}{\sqrt 2}(v_u + h_u +i a_u)\end{array}\right),\;
S= \frac{e^{i\varphi_s}}{\sqrt 2}(v_s + h_s +ia_s) .
\label{eq:Higgs_decomposition} 
\end{equation}
We have $v_d^2+v_u^2 \equiv v^2 \approx \unit[246]{GeV}$, and 
\begin{equation}
\tan\beta = \frac{v_u}{v_d} \;.
\end{equation}
The VEV of the scalar part of $\hat{S}$ dynamically generates the effective $\mu_\text{eff}$ parameter 
\beq 
\mu_{\text{eff}}= \frac{\lambda v_s e^{i\varphi_s}}{\sqrt{2}}
\eeq
through the first term in the superpotential.
The mass eigenstates $h_i$ ($i=1,...,5$) are obtained after rotating from the interaction to the mass basis. We apply two consecutive
rotations, where the first rotation ${\cal R}^G$ singles out the
would-be Goldstone boson, and the second one, ${\cal R}$, performs the
rotation to the mass eigenstates,
\begin{align}
(h_d,h_u,h_s,a,a_s, G^0)^T &=  \mathcalR^G~(
h_d,h_u,h_s,a_d,a_u,a_s)^T \nonumber \\ 
(H_1,H_2,H_3,H_4,H_5, G^0)^T &= \mathcalR ~(h_d,h_u,h_s,a,a_s,
G^0)^T\,, \label{eq:rotationtreelevel}
\end{align}
with the diagonal mass matrix\footnote{We work in the 't Hooft-Feynman
  gauge where the masses of the neutral and charged Goldstone bosons
  are equal to $Z$ and $W$ boson masses, respectively.}
\begin{align}
\diag(m_{H_1}^2,m_{H_2}^2,m_{H_3}^2,m_{H_4}^2,m_{H_5}^2,m_{G^0}^2)&=
\mathcalR \mathcalM_{hh} \mathcalR^T\,, \quad \mathcalM_{hh}=
\mathcalR^G\mathcalM_{\phi\phi}(\mathcalR^G)^T, 
\label{eq:massmatrix}
\end{align}
and 
\begin{align}
\mathcalR^G = \left( \begin{array}{cc}      1_{3\times3} & 0\\ 0 & \tilde{\cal R}^G \end{array} \right) , \quad 
\tilde{\cal R}^G =\left( \begin{array}{ccc} s_{\beta} &  c_{\beta} & 0
\\ 0 & 0 & 1 \\ c_{\beta} & - s_{\beta} &  0 \end{array}\right) \;.
\end{align}
The mass eigenstates $H_i$
are ordered by ascending mass, i.e.~$m_{H_1} \le ... \le m_{H_5}$.
The charged Higgs boson $H^-$ and
Goldstone boson $G^-$ are obtained through the rotation 
\beq 
\left( \begin{array}{c} G^-\\ H^- \end{array} \right) = \mathcalR^{G^-} \left( \begin{array}{c} h_d^-\\
h_u^- \end{array} \right), \quad  
\diag(m_{H^\pm}^2, m_{G^\pm}^2 ) =  \mathcalR^{G^-} \mathcalM_{h^+h^-}
(\mathcalR^{G^-})^T, 
\eeq
with 
\beq
\mathcalR^{G^-} = \left( \begin{array}{cc} - c_\beta & s_\beta \\
s_\beta & c_\beta \end{array} \right) \;.
\eeq
In the CP-conserving case, the CP-even and CP-odd neutral states do not mix, and the neutral Higgs sector consists of three CP-even neutral states $H_i$ ($i=1,2,3$) which are mass ordered as $m_{H_1} \le m_{H_2} \le m_{H_3}$, and two CP-odd Higgs bosons $A_j$ ($j=1,2$), mass ordered as $m_{A_1} \le m_{A_2}$. \s

After applying the minimization conditions, the chosen independent
input parameters for the tree-level NMSSM Higgs sector are, 
\beq 
 M_W, M_Z,
\alpha,\tan\beta,|\lambda|, \mu_{\text{eff}},|\kappa|,M_{H^\pm},\Re A_\kappa,\varphi_\lambda,\varphi_\kappa,\varphi_u, \varphi_s
 \,, 
\eeq
where $\alpha$ is the fine structure constant.
The remaining NMSSM input parameters (soft SUSY breaking masses and trilinear couplings) become relevant when the (required) higher-order corrections to
the Higgs boson masses are included or the supersymmetric
particle sectors of the NMSSM Lagrangian are considered. This
increases the amount of input parameters over which the scans are
performed. In \texttt{NMSSMCALC}, which we use for the generation of Higgs boson
  spectra and branching ratios, the input parameters are chosen following the SUSY Les Houches Accord \cite{Skands:2003cj}. For further details on this, cf.~also \cite{Baglio:2013iia}.

\section{Experimental Observables \label{sec:code}}

The main objective of this paper is to test our new  \texttt{NMSSMScanner} tool by deriving viable maximal di-Higgs
cross sections. For this we have to make sure that the
relevant experimental constraints are fulfilled. In the following, we
describe the observables that are tested and the computer codes that
are used. \s

{\bf Organization of the scan} \hspace*{0cm}
The utilization of different codes as well as the organization of the parameter scans is
performed with a modified version of \BSMArt
\cite{Goodsell:2023iac}. This Python code allows us to conveniently
link programs that provide predictions for a given beyond-SM (BSM)
model and to consistently pass all necessary information in the
program chain. \BSMArt can apply a variety of scanning techniques,
such as Active Learning  \cite{Goodsell:2022beo}, MCMC, random scans, or other custom algorithms to scan BSM parameter spaces.  \BSMArt was initially developed to be used in conjunction with {\tt SARAH/SPheno} \cite{Staub:2013tta,Porod:2011nf}, a multi-purpose BSM framework, and therefore was already proven to work well in a variety of different BSM scenarios \cite{Goodsell:2020rfu,Domenech:2020yjf,Goodsell:2021iwc,Goodsell:2022beo,Benakli:2022gjn,Ashanujjaman:2023etj,Darme:2023nsy,Agin:2023yoq,Faraggi:2023jzm,Agin:2024yfs}. \s

{\bf Higgs and SUSY particle spectrum and decays} \hspace*{0cm}
While {\tt SARAH/SPheno} integrates very well with \BSMArt and is able
to provide predictions for the NMSSM, we chose
to incorporate \BSMArt with the Fortran code  \NMSSMCALC
\cite{Baglio:2013iia} since it implements predictions for a larger
number of observables including the relevant
higher-order corrections with a flexible choice of renormalization
schemes. The code computes the Higgs and SUSY mass spectrum. The SUSY particle masses are calculated at leading order.  The code computes the Higgs boson mass spectrum of the CP-violating
NMSSM including the full one-loop corrections
\cite{Ender:2011qh,Graf:2012hh} and up to two-loop order in the QCD
and electroweak corrections, i.e.~the ${\cal O} (\alpha_t \alpha_s)$
\cite{Muhlleitner:2014vsa}, the ${\cal O}(\alpha_t^2)$
\cite{Dao:2019qaz} and the ${\cal O}((\alpha_t +
\alpha_\lambda+\alpha_\kappa)^2)$ \cite{Dao:2021khm}
corrections. Recently, we included the Higgs boson mass predictions for
scenarios with large SUSY mass scales \cite{Borschensky:2024utz}. 
The corrections to the $\rho $ parameter and their effect on the $W$ boson mass have been included as well in \NMSSMCALC \cite{Dao:2023kzz}. The code provides also a prediction for the leptonic anomalous magnetic moments taking into account two-loop effects \cite{Dao:2022rui} and electric dipole moment observables at one- and two-loop levels \cite{King:2015oxa} in the complex NMSSM.

Furthermore, \NMSSMCALC computes the higher-order corrections to the trilinear Higgs boson self-couplings at complete one-loop order \cite{Nhung:2013lpa} and up to two-loop ${\cal O}(\alpha_t \alpha_s)$ \cite{Muhlleitner:2015dua} and ${\cal O}(\alpha_t^2)$ \cite{Borschensky:2022pfc}. The trilinear Higgs boson self-couplings play an important role in the production of Higgs boson pairs. \s

Adapted from the Fortran code {\tt HDECAY} \cite{Djouadi:1997yw,Djouadi:2018xqq}, \NMSSMCALC also computes the Higgs boson decay widths and branching ratios, including the state-of-the-art higher-order QCD corrections and off-shell decays. In the Higgs-to-Higgs decays, we include the full one-loop corrections together with dominant two-loop corrections of ${\cal O}(\alpha_t(\alpha_s+\alpha_t))$ . They are consistently computed at the same loop order as the Higgs boson masses. Proper on-shell conditions of the Higgs bosons are ensured by taking into account the corresponding finite wave function renormalization. These corrections impact both the total widths of the Higgs bosons and their branching ratios into Higgs boson pairs. \s

Recently, \NMSSMCALC has been extended to include the computation of the 
NMSSM SUSY particle decays \cite{felixthesis}.
Adapted from the code {\tt SDECAY} \cite{Muhlleitner:2003vg}, respectively {\tt SUSY-HIT} \cite{Djouadi:2006bz}, it computes the tree-level two-body and three-body decays as well as the loop-induced decays, and it includes the next-to-leading order (NLO) SUSY-QCD corrections to decays involving colored particles. \s

{\bf Single Higgs boson signatures} \hspace*{0cm}
The loop-corrected Higgs boson masses as well as the effective Higgs boson 
couplings and/or branching ratios, that are given out by \NMSSMCALC, are subsequently used to compute the single-Higgs observables tested in experiment.  
The {\tt C++} code {\tt HiggsTools} \cite{Bahl:2022igd} computes with the effective couplings the SM-like and the non-SM-like Higgs boson production cross sections. After multiplication with the corresponding branching ratios, compatibility with both the LHC SM-like Higgs boson data and the exclusion limits from BSM Higgs boson searches at the LHC, LEP and Tevatron experiments are tested. \s

{\bf Di-Higgs Signatures}
Di-Higgs signatures can arise from non-resonant and resonant Higgs boson
pair production. Accordance with non-resonant di-Higgs searches is
validated by comparing the computed cross section for SM-like Higgs boson
pair production with the experimental results. Here, ATLAS  puts, at the 95\% CL, an upper limit
of 2.9 times the inclusive Higgs boson pair production cross section from
gluon fusion (ggF) plus vector boson fusion (VBF), $\sigma_{\text{ggF +
    VBF}}^{\text{SM}} = 32.8^{+2.1}_{-7.1}$~fb \cite{ATLAS:2024ish}. At CMS the Higgs boson pair production cross section is found to be less than 3.4 times the SM expectation at 95\% confidence level (CL), with the SM Higgs boson pair production cross section taken to be $\sigma_{\text{ggF+VBF}}^{\text{SM}} = 32.76^{+1.95}_{-6.83}$~fb \cite{CMS:2022dwd}. We use a modified version of the Fortran code {\tt HPAIR}~\cite{HPAIR} to calculate the NMSSM cross sections for Higgs boson pair production through gluon fusion into a SM-like Higgs boson pair. Developed originally for the MSSM \cite{Plehn:1996wb} it has been adapted to the NMSSM \cite{Nhung:2013lpa} and allows to include NLO QCD corrections in the heavy top limit \cite{Dawson:1998py}. 
The uncertainties of the present non-resonant Higgs boson searches as well as those expected at the High-Luminosity LHC are not expected to restrain the NMSSM parameter space. This is because the NMSSM prediction for the non-resonant production cross section of a SM-like Higgs boson pair does not differ substantially from the SM result, as single Higgs boson constraints limit deviations of the SM-like top-Higgs Yukawa coupling from the SM value to be below 10\%, and, due to SUSY relations, the allowed trilinear Higgs boson self-couplings of the SM-like Higgs boson do not differ from the SM value by more than 20\% \cite{Borschensky:2022pfc,Abouabid:2021yvw}. Due to the time-consuming calculation of the Higgs boson pair production cross section, we therefore do not perform this check during our scan. Instead, we perform it as a sanity check at the end of our program chain on the obtained parameter sample. \s

For the cross check of resonant di-Higgs search limits, {\tt HiggsTools}~\cite{Bahl:2022igd}
is applied. It multiplies the production cross sections for non-SM-like single Higgs bosons with their branching ratios into a pair of SM-like Higgs bosons and compares it with the experimental analyses. Here, {\tt HiggsTools} uses the effective couplings to compute the single Higgs boson production cross sections. In our numerical analysis, however, where we seek for the maximum resonant cross sections in various final states, we use the code \texttt{SusHi} \cite{Harlander:2012pb,Liebler:2015bka,Harlander:2016hcx} for the computation of the resonantly produced heavy Higgs bosons at next-to-next-to-leading-order (NNLO) QCD. It includes both production in gluon fusion and in association with a $b$-quark pair, where the latter does not play an important role for our scenarios, which are dominated by low $\tan\beta$ values. \s

A comment here is in order. As discussed in Ref.~\cite{Heinemeyer:2024hxa}, interference effects between resonant and non-resonant contributions (as well as loop corrections to trilinear Higgs self-couplings, cf.~e.g.~\cite{Heinemeyer:2024hxa,Arco:2025nii,Braathen:2025qxf}) can have  significant effects on the invariant mass distributions and hence on the derived exclusion limits. In the present situation, where the limits given by experiment are based on either resonant or non-resonant searches, on the theory side a decision has to be made when to apply resonant or non-resonant limits, respectively, cf.~the discussion in Ref.~\cite{Abouabid:2021yvw}. To which extent such a separation is justified depends on the importance of the interference contribution of the investigated scenario. \s

In the following, we will investigate the production of a SM-like ($H$) and a non-SM-like ($Y$) Higgs pair from the resonant production of a heavy scalar $X$, as a proof of concept of our code. Models, that lead to such signatures, comprise the possibility to resonantly enhance the SM-like Higgs pair production through two channels, the production of $X$ and the production of $Y$ and their respective subsequent decay into $HH$, depending on the mass spectrum. Since in our analysis we focus on maximizing the di-Higgs cross sections for the resonant production of a SM-like and non-SM-like Higgs pair, we will filter out scenarios where the branching ratio of the resonantly produced heavy scalar $X$ into $HY$ is maximized, so that its branching ratio in particular into a SM-like Higgs pair, BR$(X \to HH)$, is minimized. We can hence expect that the SM-like Higgs pair production proceeds dominantly non-resonantly. There is, however, still the possibility, that the resonant contribution from the $Y$ scalar with subsequent decay into $HH$ gives a significant contribution. Therefore, each benchmark point has to be investigated w.r.t.~the question to which extent  $HH$ is resonantly or non-resonantly produced, in order to take a decision on which experimental limits to apply. For the overall scan, we apply only non-resonant limits on $HH$, to save computational time. For the individual benchmark points, however, we will calculate the resonant contributions from $X$ and $Y$ production to $HH$ production (if kinematically allowed) and compare it with the result from \texttt{HPAIR}, which includes all diagrams, both resonant and non-resonant ones. This allows us to quantify the fraction of resonant contribution to Higgs pair production.  \s

{\bf SUSY particle searches} \hspace*{0cm}
The constraints from SUSY particle searches are checked with {\tt SModelS} \cite{Kraml:2013mwa,Ambrogi:2017neo,Alguero:2021dig,Altakach:2024jwk} which is already integrated into \BSMArt~\cite{Goodsell:2023iac}. 
The program calculates the required leading-order (LO) squark and gluino pair production cross sections using {\tt PYTHIA} \cite{Bierlich:2022pfr} and applies a $K$-factor for the QCD corrections obtained from {\tt NLLFast} \cite{Beenakker:1996ch,Beenakker:1997ut,Kulesza:2008jb,Kulesza:2009kq,Beenakker:2009ha,Beenakker:2010nq,Beenakker:2011fu,Beenakker:2015rna} at next-to-leading logarithmic order. 
For the computation of the electroweakino pair production cross
sections we use an in-house code and apply an approximate $K$-factor
of 1.3 for the NLO electroweak (EW) corrections. This approximation is valid as long
as the electroweakinos are lighter than the squarks
\cite{Beenakker:1996ed,Beenakker:1999xh}. The production of a mixed
squark-electroweakino pair, which in general is subdominant, is
calculated via \texttt{SModelS} and included at leading order. To
obtain the total cross section of the respective multi-particle final
states of the SUSY scenarios investigated in the individual
experimental analyses, the production cross sections are multiplied
with the SUSY particle branching ratios obtained from the new version of \texttt{SDECAY} that includes the SUSY
  particle decays in the NMSSM \cite{felixthesis}. \s

{\bf DM observables} \hspace*{0cm} 
Compatibility with the DM observables, i.e.~the measured relic density of $\Omega h^2 = 0.120 \pm 0.001$ \cite{Planck:2018vyg} and the limits from direct detection experiments \cite{Aalbers:2025LZ}, is investigated with the recently released {\tt C++} code {\tt RelExt} \cite{Capucha:2025iml}. For this purpose we extended {\tt RelExt} to the computation of the relic density and of the direct detection cross section in the NMSSM. For the comparison with the direct detection limits, the effective spin-independent DM-nucleon scattering cross section is calculated. It is obtained by multiplying the cross section with the ratio of the computed NMSSM relic density and the measured value of 0.12. In this way, possible NMSSM DM under-abundance is consistently included in the derivation of the direct detection signal for the NMSSM DM candidate.\footnote{We compared our numbers for direct detection with {\tt MicrOMEGAs} \cite{Belanger:2001fz,Belanger:2004yn,Belanger:2006is,Alguero:2023zol} and found agreement.}  \s  

{\bf Electroweak precision observables} For the check of the compatibility with the electroweak precision observables we use the $W$-boson mass prediction obtained by \NMSSMCALC \cite{Dao:2023kzz} and compare it with the world average given for the $W$ boson mass, restricting it to be within $M_W= 80.3692 \pm 0.0266 \pm 0.01$~GeV.\s

More specifically, in the scan that we performed for the results presented here, we used the following code versions:
 \texttt{BSMArt} 1.3~\cite{Goodsell:2023iac}, \texttt{LHAPDF} 6.5.3~\cite{Buckley:2014ana}, \texttt{LoopTools} 2.16~\cite{Hahn:1998yk}, \texttt{HiggsTools} 1.1.3~\cite{Bahl:2022igd} with \texttt{HBdataset} 1.6 + \texttt{HSdataset} 1.1, \texttt{SusHi} 1.7.0~\cite{Harlander:2012pb,Liebler:2015bka,Harlander:2016hcx}, \texttt{RelExt} 1.0 (NMSSM branch)~\cite{Capucha:2025iml}, and \texttt{SModels} 2.3.3~\cite{Kraml:2013mwa,Ambrogi:2017neo,Alguero:2021dig,Altakach:2024jwk}.

\section{Results \label{sec:results}}

\subsection{The Parameter Scan}
The SM input parameters are taken as 
\beq
\begin{array}{lcllcl}
G_F&=& 1.66370 \, 10^{-5} \mbox{GeV}^{-2} & \quad m_\mu &=& 0.105658367 \mbox{ GeV} \\[0.1cm]
\alpha_{\text{em}}^{-1} (M_Z) &=& 127.995 & \quad m_\tau &=& 1.77682 \mbox { GeV}\\[0.1cm]
\alpha_S (M_Z) &=& 0.1179 & \quad \overline{m}^{\text{MS}}_s (2 \mbox{ GeV} ) &=& 0.95 \mbox{ GeV}\\[0.1cm]
M_Z &=& 91.1876 \mbox{ GeV} & \quad \overline{m}^{\text{MS}}_c (\overline{m}^{\text MS}_c) &=& 1.274 \mbox{ GeV} \\[0.1cm]
M_W &=& 80.3790 \mbox { GeV} & \quad \overline{m}^{\text{MS}}_b (\overline{m}_b^{\text{MS}}) &=& 4.18 \mbox{ GeV} \\[0.1cm]
& & & \quad m_t &=& 172.76 \mbox{ GeV}.
\end{array}
\eeq
In Tab.~\ref{tab:scan}, we list the input parameters over which the scans are performed together with their respective scan boundaries. In this first presentation of our results, we resort to the CP-conserving NMSSM, so that all input  parameters are taken real. In accordance with the SUSY Les Houches Accord (SLHA) format,  the soft SUSY breaking masses and trilinear couplings, the  higgsino mass and $\tan\beta$  are understood as $\overline{\mbox{DR}}$ parameters at the scale $\mu_0 = M_{\text{SUSY}}= \sqrt{m_{\tilde{Q}_3} m_{\tilde{t}_R}}$, which is also the renormalization scale used in the computation of the higher-order corrections. The charged Higgs boson  mass is taken as input parameter and has been chosen to be larger than 600~GeV in order to account for the type-II $B$ physics constraint from $b \to s \gamma$  \cite{Deschamps:2009rh,Mahmoudi:2009zx,Hermann:2012fc,Misiak:2015xwa,Misiak:2017bgg,Misiak:2020vlo}. Consequently, all other doublet-like non-SM-Higgs boson masses will also be rather heavy. To account for the perturbative unitarity limit we furthermore apply the rough constraint of \cite{King:2012tr} 
\beq
\lambda^2 + \kappa^2 \leq 0.7 \;.
\eeq
\begin{table}[h]
\centering
\begin{minipage}[t]{0.5\textwidth}
\centering
    \begin{tabular}[t]{ll}
        parameter              & scan range [TeV]                \\ \hline
$m_{H^\pm}$            & [0.6, 4]               \\
$M_1,M_2$              & [0.1, 4]                    \\
$M_3$                  & [0.4, 4]               \\
$m_{\tilde{Q}_3}, m_{\tilde{t}_R}$, $m_{\tilde{b}_R}$      & [0.4, 4]               \\
$m_{\tilde{L}_3}, m_{\tilde{\tau}_R}$       & [0.4, 4]               \\
$m_{\tilde{Q}_2}, m_{\tilde{u}_R}$       & [0.4, 4]               \\
\end{tabular}
\end{minipage}\hfill
\begin{minipage}[t]{0.5\textwidth}
\centering
    \begin{tabular}[t]{ll}
parameter              & scan range                 \\ \hline
$\mu_{\text{eff}}$               & [0.1, 4] TeV             \\
$A_{t,b,\tau}$                  & [$-4$, 4] TeV                 \\
$A_{\kappa}$           & [-4, 0.1] TeV \\
$\tan\beta$            & [1, 20]                    \\
$\lambda$              & [0.01, 1]                 \\
$\kappa$               & [0.01, 1] \\     
\end{tabular}
\end{minipage}
\caption{Ranges for the scan over the NMSSM parameter
  space. We set $m_{\tilde{L}_1}=m_{\tilde{L}_2}=m_{\tilde{Q}_1}=m_{\tilde{Q}_2}$ and $m_{\tilde{u}_R}=m_{\tilde{c}_R}=m_{\tilde{d}_R}=m_{\tilde{s}_R}=m_{\tilde{e}_R}=m_{\tilde{\mu}_R}$.} 
\label{tab:scan}
\end{table}
Consistency with the experimental Higgs boson results requires one of the neutral Higgs bosons to have a mass of 125~GeV and behave very SM-like. The latter implies that the mass eigenstate, which we will call from now on $H$, has a large $h_u$ component. For our scan we demand the loop-corrected\footnote{ Note that, in contrast to \cref{sec:model}, here and in the following we always refer to loop-corrected mass values for all neutral Higgs bosons.} mass of this SM-like Higgs boson to lie in the range
\beq
124  \le m_H \le 126 \mbox{ GeV}
\eeq
at ${\cal O}(\alpha_t (\alpha_s + \alpha_t))$ in the default mixed $\overline{\mbox{DR}}$-OS scheme introduced in \cite{Graf:2012hh} and with $\overline{\mbox{DR}}$ renormalization in the top/stop sector. In order to be consistent with the loop order used in the fixed-order prediction of the loop-corrected trilinear Higgs boson self-couplings and in order to take into account mixing effects between the singlet field and the SM-like doublet field, we intentionally do not include the two-loop ${\cal O}((\alpha_t+\alpha_\lambda+\alpha_\kappa)^2+\alpha_t \alpha_s)$ corrections\footnote{This loop order is available for the Higgs boson masses, but not yet available for the trilinear self-couplings.} or use the hybrid-effective-field-theory (EFT) Higgs boson mass prediction \cite{Borschensky:2024utz}. However, we explicitly checked that for all of our valid parameter points obtained in the scan, the hybrid-EFT Higgs boson mass prediction stays within a 2-3~GeV interval of the measured value. \s

In this first presentation of sample results obtained with the new package \texttt{NMSSMScanner}, we optimized an MCMC scan by using appropriate likelihood functions for each individual final state considered. We generated seed points for the MCMC using a traditional uniform random scan within the scan ranges defined in Tab.~\ref{tab:scan}. Subsequently, in order to obtain the maximum cross section values for the resonant production of a Higgs boson $X$ which then decays into a SM-like plus non-SM-like Higgs boson pair $HY$, we performed MCMC scans using likelihood functions $\mathcal{L}$ optimized for our needs, within predefined mass grids (along the experimental analyses) with the starting points given by the random points.
In the case of a scalar resonance, we chose the likelihood function
\beq
\mathcal{L}_{\text{max}}^s = \exp{(s_0 (\sigma \times \mathrm{BR}) \slash \mu )} \,,
\eeq
where the normalization $\mu$ is given by $\mu=\sigma_\text{seed}\times\mbox{BR}_\text{seed}$ calculated from the starting seed point. Here, $\sigma$ denotes the cross section for the resonant $HY$ production and BR stands for the product of the branching ratios of $H$ and $Y$ into the considered final state.
We set $s_0=2$ which was found to be optimal for the convergence of the scan. 
For the maximization of the case where we have a pseudoscalar resonance $X$ and the lighter pseudoscalar $Y$ decays into a photon pair, we define a different likelihood function to enhance the efficiency of the scan. Since the maximization of this channel sensitively depends on the singlet admixture to the lighter pseudoscalar $A_1$, which we denote in the following
as $\mathcal{R}_{A_1 a_s}^2$, we applied the likelihood 
\begin{equation}
    \mathcal{L}_{\text{max}}^p = \mathcal{L}_{\text{max}}^s \times \exp (s_0\, \mathcal{R}_{A_1 a_s}^2 \slash \mu_s)\,,
\label{eq:likelihoodsinglet}
\end{equation}
where again $s_0=2$  and $\mu_s$ is the $\mathcal{R}_{A_1s}^2$ value of the starting seed point. 
The usage of this likelihood function did not drastically change the result, however.

\subsection{Scan Results}
In the following, we present our results for the maximum cross sections obtained for the production of a SM-like Higgs boson ($H$) together with a non-SM-like one ($Y$), produced from the resonant decay of a heavier Higgs boson ($X$), which is produced in gluon fusion at the LHC at a center-of-mass energy of 13~TeV,
\begin{equation}
\sigma^{s/p}_{HY} = \sigma(gg \to X) \times \mbox{BR} (X \to HY) \;. \label{eq:dihiggsprocess}
\end{equation}
Since we focus on the CP-conserving NMSSM, the Higgs boson spectrum consists of CP eigenstates given by three CP-even Higgs bosons $H_{1,2,3}$ and two CP-odd Higgs bosons $A_{1,2}$ as well as the charged Higgs bosons $H^\pm$. 
In the process Eq.~(\ref{eq:dihiggsprocess}), the heavier ($X$) can here be either scalar ($s \equiv H_3$) or pseudoscalar ($p \equiv A_2$). We hence have
\beq
\sigma_{HY}^s &=& \sigma (gg \to H_3) \times \mbox{BR} (H_3 \to H_1 H_2) \\
\sigma_{HY}^p &=& \sigma (gg\to A_2) \times \mbox{BR} (A_2 \to H_{1/2} A_1) \;.
\eeq
Depending on the specific benchmark scenario, the SM-like Higgs boson $H$  can be either the lightest ($H_1$) or the next-to-lightest ($H_2$) scalar Higgs boson. 
We remind the reader that the $X$ production cross section is calculated at NNLO QCD with the help of \texttt{SusHi} 1.7.0\cite{Harlander:2012pb,Liebler:2015bka,Harlander:2016hcx}. In the following, we only present parameter sets obtained from our scan that respect the above speci\-fied constraints. 
We give results for the mass pairs ($m_X,m_Y$) orienting ourselves along the experimental searches. 
The lower mass limit for $m_X$ in our scan sample is 600~GeV as result of the applied $B$ physics constraint that restricts the charged Higgs boson mass (and consequently all heavy doublet states) to be above 600~GeV. We note that this constraint is very conservative and that charged Higgs boson masses of about \unit[400]{GeV}, and hence also $m_X$ masses, are in principle still possible if the supersymmetric contributions to e.g. $b\to s\gamma$ are taken into account. In the following, we present the results for the following final states,
\begin{eqnarray}
\begin{array}{lll}
(b\bar{b})(b\bar{b}) \mbox{ final state:} & \sigma_{HY}^{s/p}\times\mbox{BR}(H\to
b\bar{b})\times \mbox{BR} (Y \to b\bar{b}) & \mbox{ \cite{CMS:2022suh,CMS:2024bds} }\\[0.1cm]
 (b\bar{b})(\tau\tau) \mbox{ final state}: & \sigma_{HY}^{s/p}\times \mbox{BR}(H\to \tau^+ \tau^-)\times \mbox{BR} (Y \to b\bar{b}) & \mbox{ \cite{CMS:2021yci} }\\[0.1cm]
 (b\bar{b})(\gamma\gamma) \mbox{ final state}: &\sigma_{HY}^{s/p}\times \mbox{BR}(H\to \gamma\gamma)\times \mbox{BR} (Y \to b\bar{b}) & \mbox{ \cite{CMS:2023boe,ATLAS:2024auw,ATLAS:2025nda} }\\
& \sigma_{HY}^{s/p}\times \mbox{BR}(H\to b\bar{b})\times \mbox{BR} (Y \to \gamma\gamma) & \mbox{ \cite{CMS:2025qit} }\\[0.1cm]
(\tau\tau)(\gamma\gamma) \mbox{ final state}: & \sigma_{HY}^{s/p}\times \mbox{BR}(H\to \gamma\gamma)\times \mbox{BR} (Y \to \tau\tau)  & \mbox{ \cite{CMS:2025tqi}}  \\
& \sigma_{HY}^{s/p}\times \mbox{BR}(H\to \tau\tau)\times \mbox{BR} (Y \to \gamma\gamma) & \mbox{ \cite{CMS:2025tqi} } \\[0.1cm]
(b\bar{b}) (t\bar{t}) \mbox{ final state}: & 
\sigma_{HY}^{s/p}\times \mbox{BR}(H\to b\bar{b})\times \mbox{BR} (Y \to t\bar{t}) \\[0.1cm]
(6b) \mbox{ final state from } (3H): &
\sigma_{HY}^{s}\times\mbox{BR} (Y \to HH) \times \left(\mbox{BR}(H\to b\bar{b})\right)^3 & \mbox{\cite{ATLAS:2024xcs,CMS:2025gos}}\\[0.1cm]
(\gamma\gamma)(WW) \mbox{ final state}: &
\sigma_{HY}^{s}\times \mbox{BR}(H\to \gamma\gamma)\times \mbox{BR} (Y \to WW) & \mbox{ \cite{ATLAS:2024xkk} } \\
\end{array}
\label{eq:channels}
\end{eqnarray}
The citations refer to the ATLAS and CMS papers, where these final
states have been studied. 
Further benchmark points for different mass pair values and different decay channels can be generated on request.  

\subsubsection{Maximum Cross Section Values}
\begin{figure}[t!]
\begin{center}
\includegraphics[width = 7.4cm]{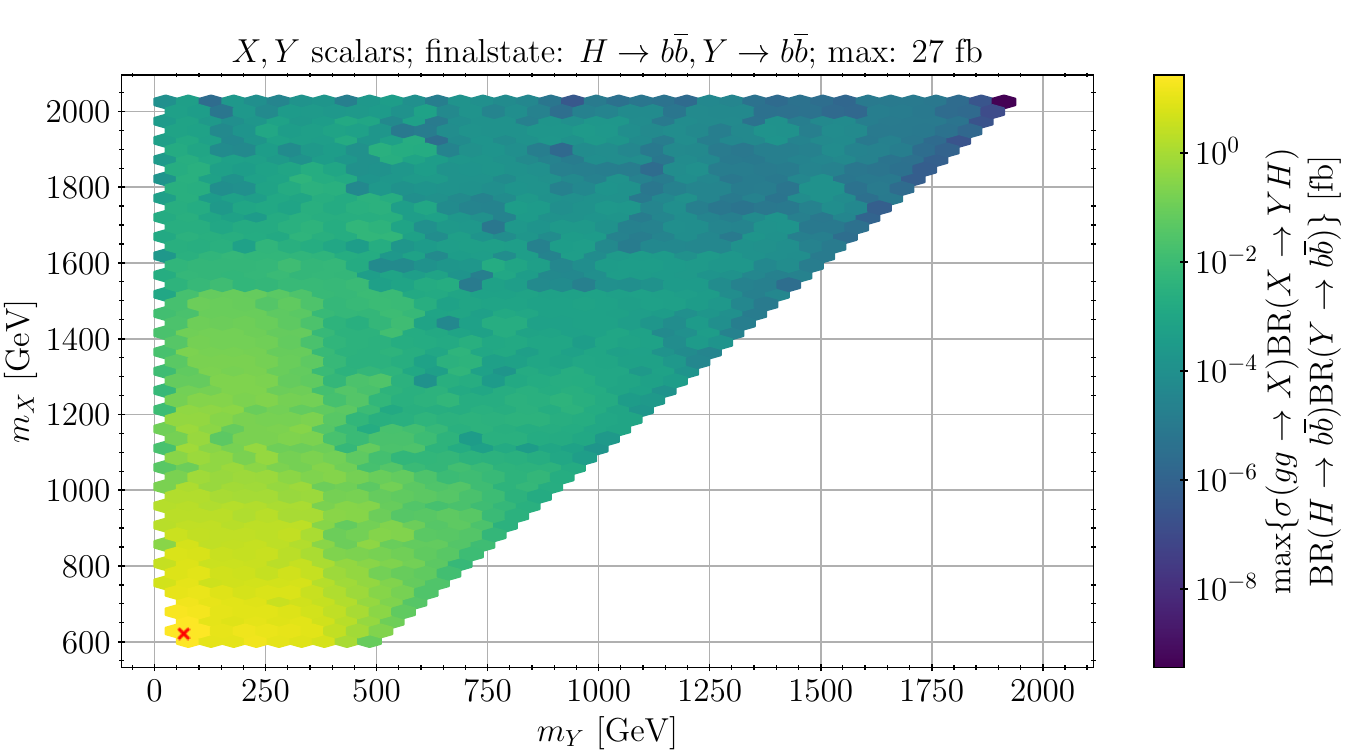}
\includegraphics[width = 7.4cm]{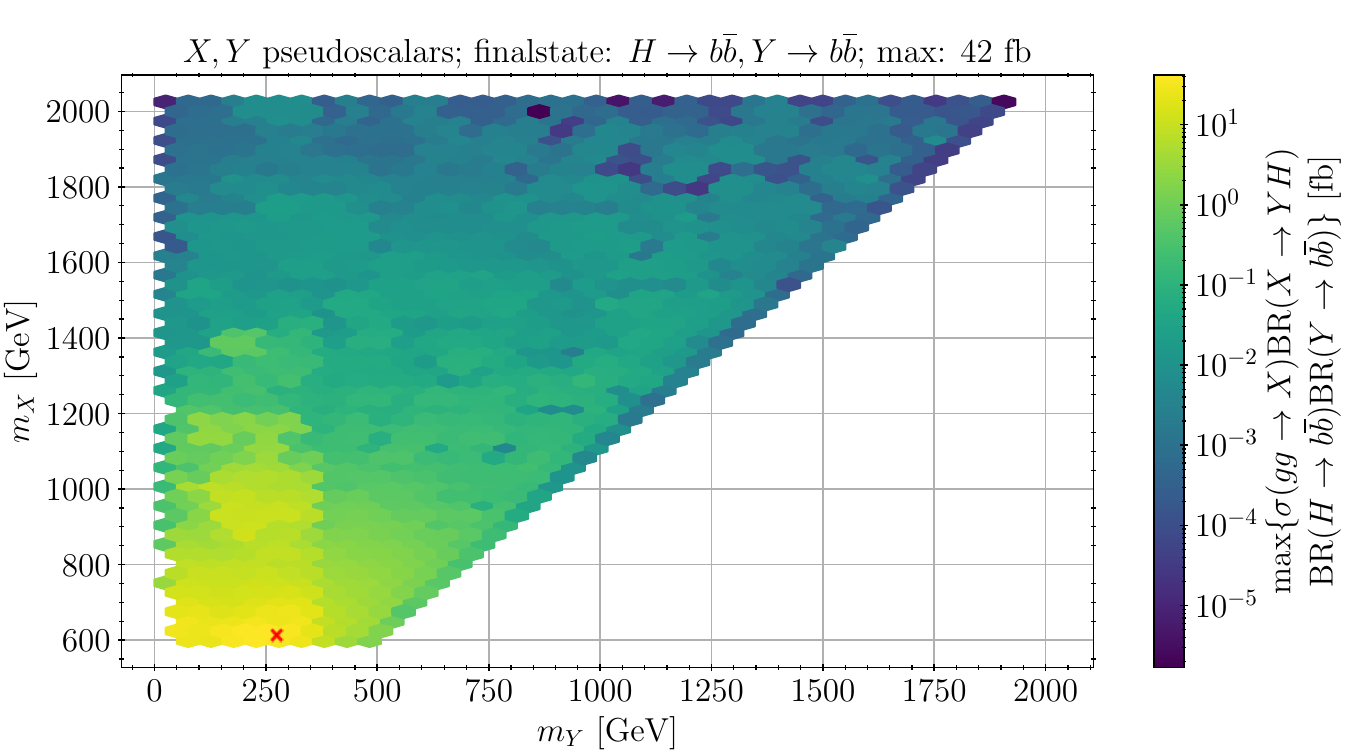}
\caption{Maximal  cross section values in fb in the $m_X$-$m_Y$ plane for the resonant production of a SM-like and non-SM-like Higgs boson pair from a scalar (left) and pseudoscalar (right), with both final state Higgs bosons decaying into a $b$-quark pair. The red crosses mark the benchmark point with the maximum cross section value. 
\label{fig:max4bproduction}}
\end{center}
\end{figure}

In the following we show exemplary results for the $4b$ final state
from resonant production of a mixed di-Higgs pair consisting of a
SM-like and non-SM-like Higgs boson subsequently decaying into a $b\bar{b}$
final state each. The results of the measurement of these processes
have been presented by CMS in
\cite{CMS:2022suh}. Figure~\ref{fig:max4bproduction} shows as color
bar the maximum cross section values in the $(m_X,m_Y)$ plane using a hexagonal binning. For these "hexagon" plots we group together parameter points lying in one hexagon of 50~GeV size both in the $m_X$ and $m_Y$ direction, and out
of these points select the one leading to the largest cross section
value, which is then plotted, respectively, shown as color bar. This 
choice is justified by the fact that, in the NMSSM, masses are not 
input quantities but computed from the model parameters. The
plot on the left (right) shows the production of a SM-like Higgs boson $H$ and a scalar (pseudoscalar) non-SM-like $Y$ from the decay of a resonantly produced scalar (pseudoscalar) $X$. The benchmark points that lead
to the maximum cross section values in this final state are marked by
a cross. \s

In the following we will give the results of the thus obtained maximum
cross section values for various final states, that have been
investigated by the LHC experiments. Further cross section values for
specific $(m_X,m_Y)$ combinations in these or different final states
can be provided on request. In Tab.~\ref{tab:scalarres}, we give the
maximum cross section values for the $4b$, the $(2b)(2\tau)$, the
$(2b)(2\gamma)$, and the $(2\tau)(2\gamma)$ final states for a
resonantly produced scalar, $X\equiv$ scalar, together with the
benchmark point names and the corresponding tables given in the appendix, that list their
input parameters and relevant information on single- and di-Higgs
cross sections. The corresponding table for the resonant production of
a pseudoscalar, $X\equiv$ pseudoscalar, is given by Tab.~\ref{tab:pseudoscalarres}. \s

\renewcommand{\arraystretch}{1.3}
\begin{table}[t]
\begin{center}
\begin{tabular}{|c||c||c||c|}
\hline
\texttt{BP} name & \texttt{BPs4b}, Tabs.~\ref{tab:bps4b}, \ref{tab:valuesbps4b}
  &  \texttt{BPs2b2gam}
    Tabs.~\ref{tab:bps2b2gam}, \ref{tab:valuesbps2b2gam} &
\texttt{BPs2gam2b}, Tabs.~\ref{tab:bps2gam2b},
    \ref{tab:valuesbps2gam2b} \\ \hline \hline
  final state & $(H\to b\bar{b})(Y \to b\bar{b})$ &
$(H\to b\bar{b})(Y \to \gamma\gamma)$ &
 $(H\to \gamma\gamma)(Y \to b\bar{b})$   \\
 $\sigma^s_{\text{max}}$ [fb]  & 27 & 0.119 & 0.121 \\ \hline   
final state & $(H\to \tau\bar\tau)(Y \to b\bar{b})$ & $(H\to \tau\bar\tau)(Y
\to \gamma\gamma)$& $(H\to \gamma\gamma)(Y \to \tau\bar\tau)$   \\
  $\sigma^s_{\text{max}}$ [fb]  & 2.9 & 0.013 & 0.012 \\ \hline 
\end{tabular}
\caption{Benchmark points for the maximum cross section values for resonant scalar production
  $X\equiv H_3$, decaying into a SM-like Higgs boson $H$ and a non-SM-like
  scalar Higgs boson  $Y=H_1$ or $H_2$ in the $4b$, the $(2b)(2\tau)$, the
$(2b)(2\gamma)$, and the $(2\tau)(2\gamma)$ final states, together
with the reference to the corresponding tables containing all relevant
information. \label{tab:scalarres}}
\end{center}
\end{table}

\renewcommand{\arraystretch}{1.3}
\begin{table}[t]
\begin{center}
\begin{tabular}{|c||c||c|}
\hline
\texttt{BP} name & \texttt{BPp4b}, Tabs.~\ref{tab:bpp4b}, \ref{tab:valuesbpp4b}
  &  \texttt{BPp2b2gam}
    Tabs.~\ref{tab:bpp2b2gam}, \ref{tab:valuesbpp2b2gam}  \\ \hline\hline
  final state & $(H\to b\bar{b})(Y \to b\bar{b})$ & \\
  $\sigma^p_{\text{max}}$ [fb]  & 42 & \\ \hline
  final state & $ (H\to \tau\tau )(Y \to b\bar{b})$ &  \\
 $\sigma^p_{\text{max}}$ [fb]  & 4.5 & \\ \hline 
final state & $(H\to \gamma \gamma)(Y \to b\bar{b})$ & $(H\to b\bar{b})(Y
\to \gamma\gamma)$  \\
  $\sigma^p_{\text{max}}$ [fb]  & 0.16 &  0.35  \\ \hline
final state &   $(H\to \gamma \gamma)(Y \to \tau\tau)$ & $ (H\to \tau\tau)(Y \to
\gamma\gamma)$ \\
  $\sigma^p_{\text{max}}$ [fb]  & 0.02 & 0.038 \\ \hline
\end{tabular}
\caption{Benchmark points for the maximum cross section values for resonant
  pseudoscalar production $X \equiv A_2$, decaying into a SM-like
  Higgs boson $H$ and a pseudoscalar $Y=A_1$ in the $4b$, the $(2b)(2\tau)$, the $(2b)(2\gamma)$, and the $(2\tau)(2\gamma)$ final states, together with the reference to the corresponding tables containing all relevant information. \label{tab:pseudoscalarres}}
\end{center}
\end{table}

We remark that the benchmark points \texttt{BPp4b} and \texttt{BPp2b2gam} feature the possibility of having measurable $4b$ final state rates from the production of two non-SM-like Higgs bosons, namely the $A_1 A_1$ production from a heavy resonant scalar $H_3$. More specifically we have
\begin{eqnarray}
\begin{array}{ll}
\mbox{\underline{\texttt{BPp4b}:}} & \sigma(H_3)^{\text{NNLO}} = 31\mbox{ fb}, \, \mbox{BR} (H_3 \to A_1 A_1) = 0.66, \; \mbox{BR} (A_1 \to b\bar{b}) = 0.74 \\
& \sigma^{\texttt{NNLO}}\times \mbox{BR}(H_3 \to A_1 A_1) \times (\mbox{BR}(A_1 \to b\bar{b}))^2 = 11 \mbox{ fb} \;.
\end{array}
\end{eqnarray}
and
\begin{eqnarray}
\begin{array}{ll}
\mbox{\underline{\texttt{BPp2b2gam}:}} & \sigma(H_3)^{\text{NNLO}} = 109\mbox{ fb}, \, \mbox{BR} (H_3 \to A_1 A_1) = 0.49, \; \mbox{BR} (A_1 \to b\bar{b}) = 0.51 \\
& \sigma^{\texttt{NNLO}}\times \mbox{BR}(H_3 \to A_1 A_1) \times (\mbox{BR}(A_1 \to b\bar{b}))^2 = 14 \mbox{ fb} \;.
\end{array}
\end{eqnarray}

\renewcommand{\arraystretch}{1.3}
\begin{table}[h!]
\begin{center}
\begin{tabular}{|c||c|c|}
\hline
\texttt{BP} name & \texttt{BPs2b2t}, Tabs.~\ref{tab:bps2b2t}, \ref{tab:valuesbps2b2t}
  &  \texttt{BPp2b2t}
    Tabs.~\ref{tab:bpp2b2t}, \ref{tab:valuesbpp2b2t}  \\ \hline\hline
  final state & $(H\to b\bar{b})(Y \to t\bar{t})$ & $(H\to b\bar{b})(Y \to t\bar{t})$ \\
  $\sigma^{s/p}_{\text{max}}$ [fb]  & 30 & 37 \\ \hline \hline \hline
\texttt{BP} name & \texttt{BPs3H6b}, Tabs.~\ref{tab:bps6b}, \ref{tab:valuesbps6b}
  &  \texttt{BPs2gamma2w}
    Tabs.~\ref{tab:bps2gam2w}, \ref{tab:valuesbps2gam2w}  \\ \hline\hline  
  final state & $(H\to b\bar{b}) (Y \to HH \to 4b)$ & $ (H\to \gamma\gamma)(Y \to WW)$ \\
 $\sigma^{s}_{\text{max}}$ [fb]  & 4.03 & 0.104 \\ \hline 
\end{tabular}
\caption{Benchmark points for the maximum cross section values into heavier final states: 
Upper: $HY$ production from resonant scalar $H_3$ (left) and pseudoscalar $A_2$ (right) production with $HY$ decaying into $(b\bar{b})(t\bar{t})$ final states. Lower: $HY$ production from resonant scalar production $H_3$ with $HY$ decaying into $(H\to b\bar{b}) (Y\to HH \to 4b)$ (left) and into $(\gamma\gamma)(WW)$ (right). \label{tab:heavier}}
\end{center}
\end{table}

In Tab.~\ref{tab:heavier}, we present benchmark points where the non-SM-like Higgs bosons in the final state decays into heavier final states, i.e.~$t\bar{t}$, SM-like Higgs boson $HH$ and $WW$ final states. \s

As outlined above, for all presented benchmark points we computed the fraction of the resonant contribution to SM-like $HH$ production. The resonant contribution to $HH$ production is found to be below 15\% for all of them, with the exception of the benchmark point \texttt{BPp2b2gam}, on which we will comment below. For the other benchmark points, the application of non-resonant search limits should be safe (see our discussion above). To corroborate this, a dedicated analysis by the experimental collaborations is required, however, which is beyond the scope of this work. We also note, that in the $6b$ final state some points from our scan had to be excluded. While their respective Higgs coupling to the top-quarks and the trilinear Higgs self-coupling are very SM-like, the contribution from the resonantly produced $Y$ with subsequent decay into $HH$ enhances the $HH$ cross section beyond the upper limit on the $HH$ production cross section, which is given by experiment to be 2.5 the SM value at 95\% C.L.~\cite{CMS:2026nuu}.

\subsubsection{Discussion}
\paragraph{Light final states $b\bar{b}$, $\tau\tau$, $\gamma\gamma$:} We start by discussing the benchmarks for final state Higgs boson decays into lighter final states, cf.~Tabs.~\ref{tab:scalarres} and \ref{tab:pseudoscalarres}. 
As can be inferred from the tables describing the benchmark points,
the overall Higgs boson spectrum is rather light. The mass of the resonantly
produced Higgs boson $X$ takes values at the lower scan
boundary, i.e.~has a mass around 600~GeV. \s

In case of a heavy scalar resonance, $X \equiv H_3$, both lighter scalar Higgs bosons $H_1$ and $H_2$
have rather low masses, and the SM-like Higgs boson $H$ can be the lightest
or the next-to-lightest Higgs boson, $H=H_1$ or $H_2$ depending on the benchmark
point. In these scenarios, the heavier pseudoscalar $A_2$ and the
non-SM-like lighter scalar, $Y=H_1$ or $H_2$ depending on the benchmark point,
is singlet-like. The SM-like Higgs boson $H$, the heavier scalar $H_3$, and
the lighter pseudoscalar $A_1$ are doublet-like, with $A_1$ being
close in mass to $H_3$ with a mass value around 600~GeV. The $A_2$ masses
range between 700 and 900~GeV. \s 

For a heavy pseudoscalar resonance, $X \equiv A_2$, the SM-like
Higgs boson is always the lightest scalar, $H = H_1$. The resonant $A_2$
mass is around 600~GeV. The $A_2$ is doublet-like as well as the
$H_2$ which is close in mass to $A_2$. The $H_3$ and $A_1$ are
singlet-like. The lighter pseudoscalar mass is heavier than the SM-like
mass with mass values above 240~GeV. The singlet-like
heavier scalar $H_3$ has masses not much above $H_2$, below 700~GeV in
these maximum cross section scenarios. \s

In all scenarios, due to supersymmetry, the charged Higgs boson and the
doublet-like non-SM-like Higgs bosons are close in mass with mass values around
600~GeV. 
The total widths are at most about 20~GeV and for the
singlet-like as well as the SM-like Higgs boson, the total widths are rather
small compared to the masses such that the narrow-width approximation, which was applied here, is well-motivated. The values of $\tan\beta$ are small, as is usual for NMSSM
scenarios. The NMSSM-specific couplings $\lambda$ and $\kappa$ range around 0.5 as
consequence of the imposed rough unitarity bound. The soft-SUSY
breaking stop parameter is rather large with values around -3 to
-4~TeV as a consequence of the applied constraint on the SM-like Higgs boson mass. \s

The results show that, in case of a resonant pseudoscalar $X=A_2$, the
cross sections are larger than for a resonant scalar when comparing
the corresponding final states. This is due to the larger gluon fusion
cross sections for pseudoscalar production. The rates for $4b$ final
states reach several tens of fb. The $(2b)(2\tau)$ production is about a
factor 10 smaller and reaches a few fb, which should still be
measurable. The $(2b)(2\gamma)$ final states have rates of a few
tenths of fb, which will be a challenge, but profits from the photons
in the final states. The $(2\tau)(2\gamma)$ final states are another
factor of 10 reduced. \s

Finally, let us comment on \texttt{BPp2b2gam}. Contrary to all other presented benchmark points into light final states, here the resonant contribution from $Y$ production with subsequent decay into $HH$ amounts to 32\% of the total Higgs pair production cross section (which includes both non-resonant and resonant diagrams). While the non-resonant search limits do not exclude this benchmark point, the assumption of applying non-resonant search limits to check for the validity of this point may hence still be questionable. This requires a closer investigation, that is far beyond the scope of this paper. We want to make aware of it, however, that this benchmark point has to be taken with a grain of salt.

\paragraph{Heavier final states $t\bar{t}$, $WW$, $H_1 H_1$:} We now discuss the benchmark scenarios where the heavier non-SM-like Higgs boson in the final state decays into heavier particles, cf.~Tab.~\ref{tab:heavier}. The cross sections into the $(b\bar{b})(t\bar{t})$ final states (first row of Tab.~\ref{tab:heavier}) can be rather large, as the decay of the scalar/pseudoscalar Higgs boson (here $H_2$ and $A_1$, respectively)  into top-quark pairs often constitutes the main branching ratio once the kinematic threshold for the decay into top-quarks is reached. While for the pseudoscalar resonance $A_2$ the branching ratio into $H_1 A_1$ is much smaller compared to the branching ratio of the scalar resonance $H_3$ into $H_1 H_2$, its production cross section largely exceeds the $H_3$ production cross section, so that overall the $(b\bar{b})(t\bar{t})$ cross section for the pseudoscalar resonance is larger than the one for the scalar resonance with 37~fb versus 30~fb. \s

In the scenario of \texttt{BPs3H6b} the mass of the non-SM-like scalar $H_2$ is below the $t\bar{t}$ but above the $H_1H_1$ threshold such that its main branching ratio is given by the decay $H_2 \to H_1 H_1$, amounting to BR$(H_2 \to H_1 H_1)=0.435$. 
Although the branching ratio of $H_3 \to H_1 H_2$ with BR$(H_3\to H_1 H_2)=0.11$ is not very large, the large 
resonant $H_3$ production 373.77~fb leads to a rather large $3H_1$ cross section of 17.9~fb, resulting finally in a $6b$ final state cross section of 4.03~fb. \s

In case of the benchmark point \texttt{BPs2gamma2w}, stemming from the resonant $H_3$ production decaying into $H_1H_2$, the mass of $H_2$ is above the $WW$ but below the $H_1H_1$ and $t\bar{t}$ thresholds so that the dominant branching ratio is into $WW$ with a value of 0.93. The next important decays are into $ZZ$ followed by the decay into $b\bar{b}$ which is substantially less important, however. The cross section for $H_3$ production is rather large with 479~fb resulting, despite the small branching ratio into $H_1H_2$ in significant $H_1H_2$ production with 42~fb. With this, the final $(\gamma\gamma) (t\bar{t})$ state amounts to 0.1~fb. \s

In all presented benchmark points for heavier final states the total widths of the Higgs particles remain below 5\% compared to the respective mass, so that the application of the narrow-width approximation is justified. 

\begin{figure}[ht]
    \centering
\includegraphics[width=0.95\linewidth]{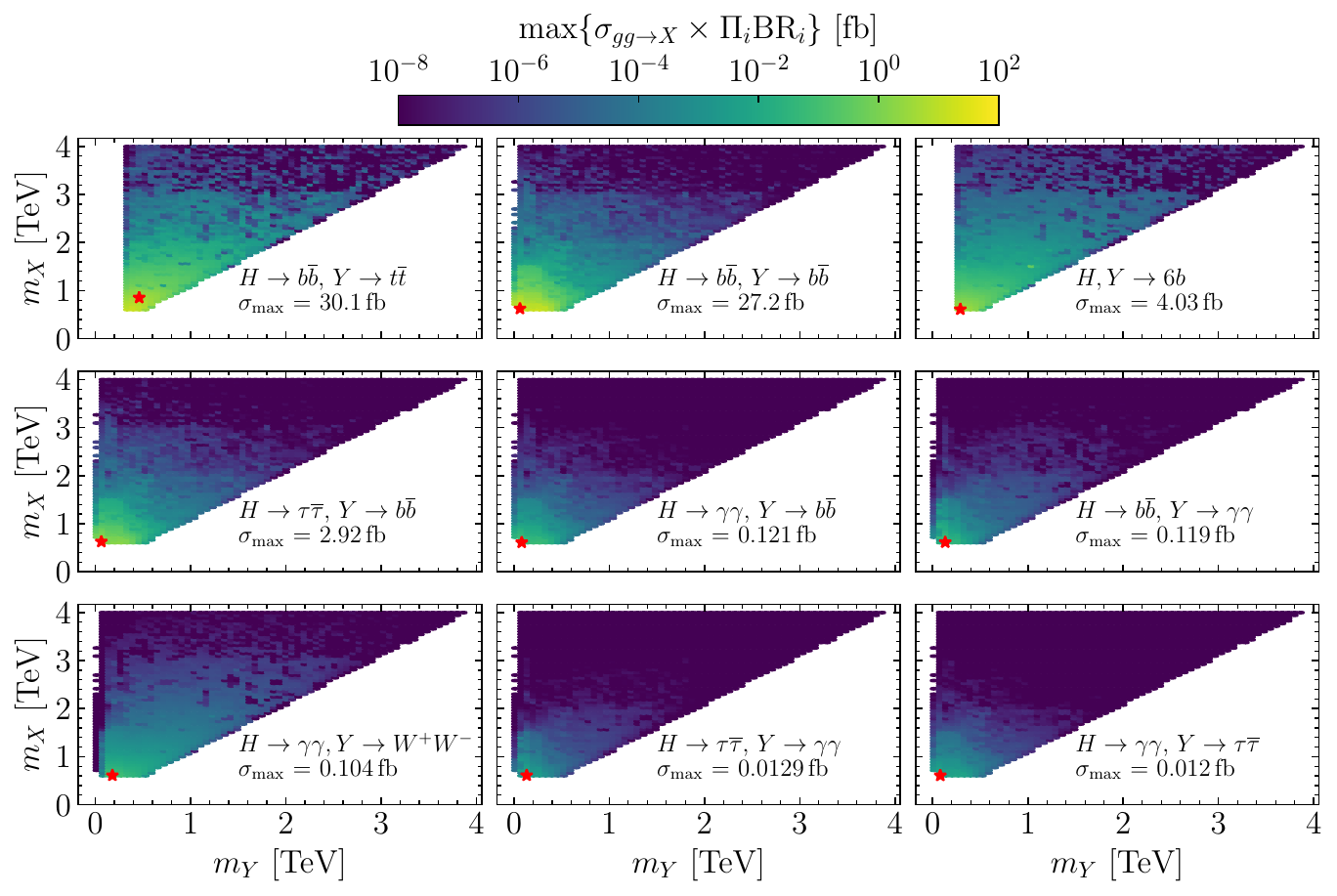}
    \caption{
    Maximal  cross section values in fb in the $m_X$-$m_Y$ plane for the resonant production of a SM-like and non-SM-like Higgs boson pair from a scalar resonance, for all decay channels considered in this work.}
    \label{fig:xsgrid}
\end{figure}
\subsection{Dominant Channels}
Figure~\ref{fig:xsgrid} shows the hexagon plots for the case of the scalar resonance in the $m_X-m_Y$ plane for all considered nine final states (given in Tab.~\ref{tab:scalarres} and in Tab.~\ref{tab:heavier} upper left and in the lower row). They are ordered by the size of the maximum cross section value that can be obtained. As can be inferred from the individual plots the largest cross sections appear in the lower left corners, i.e.~for small $X$ and $Y$ masses, as then the $X$ resonant production cross section is largest and the $s$-channel suppression of the $Y$ state is minimal. The largest cross section values are obtained for the $(b\bar{b})(t\bar{t})$ and the $(b\bar{b})(b\bar{b})$ final states. But also the $(6b)$ final state from the production of three SM-like Higgs bosons leads to cross section values of up to almost 5~fb which should be accessible. \s 

\begin{figure}
    \centering
\includegraphics[width=0.47\linewidth]{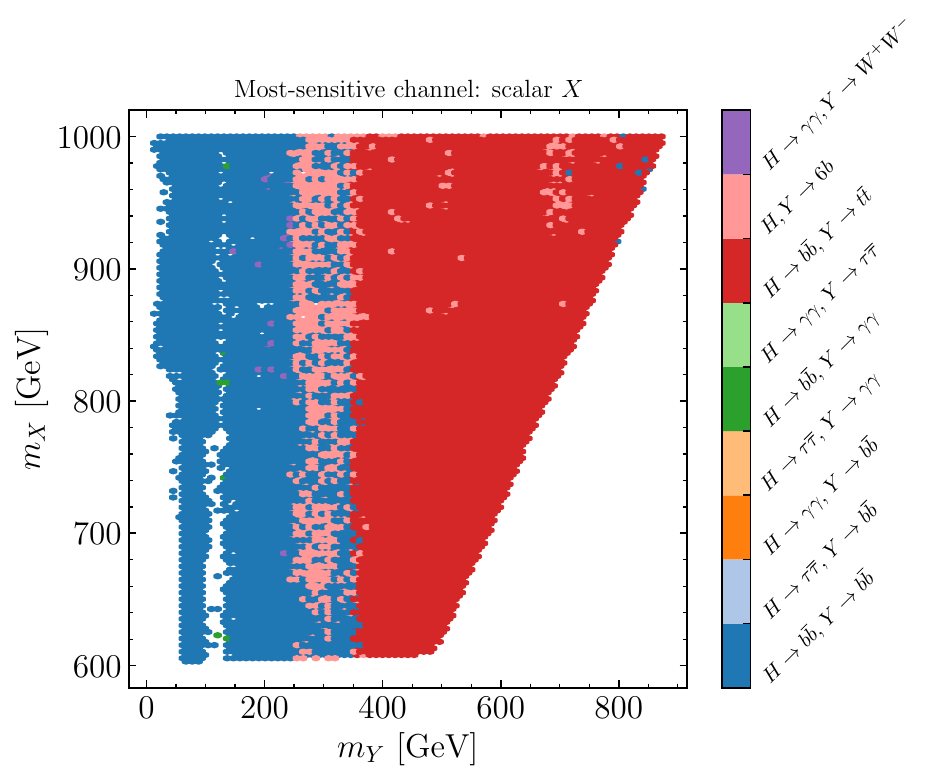}
 \includegraphics[width=0.45\linewidth]{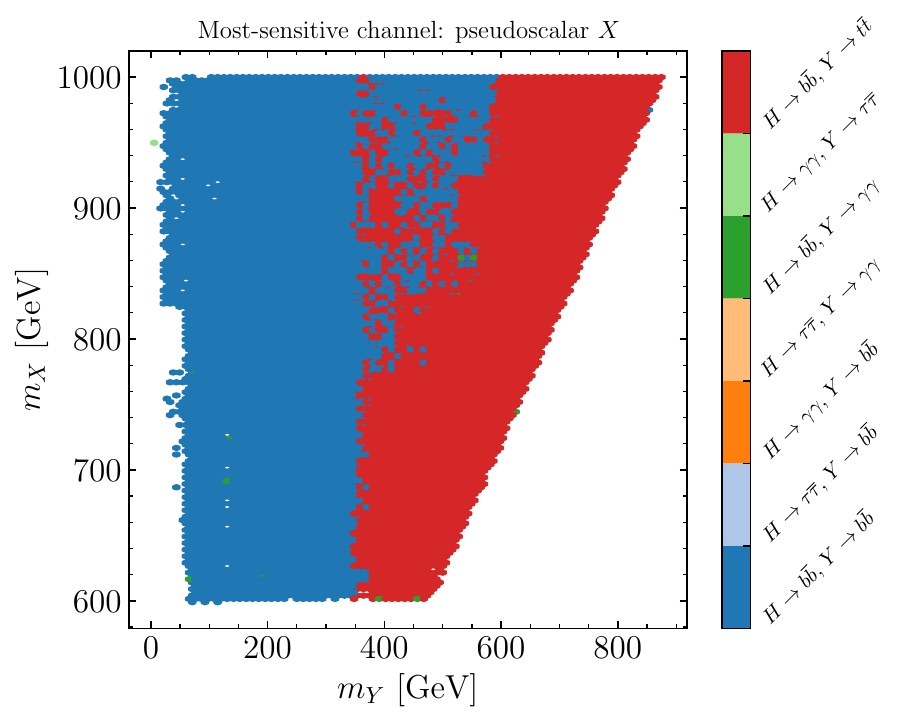}
    \caption{Dominating maximum cross section, with the final state indicated by the color bar, in the $m_X-m_Y$ plane for resonant scalar (left) and pseudoscalar (right) production.}
    \label{fig:maxchannels}
\end{figure}

In Fig.~\ref{fig:maxchannels}, we show which of the considered final states dominates for each grid mass point $(m_X,m_Y)$ in the production of the resonant scalar (left) and the resonant pseudoscalar (right). We restrict the plot range to $m_X, \, m_Y$ values below 1~TeV since for mass ranges above the di-top threshold, no additional effects compared to the ones discussed in the following were found. In both cases, scalar and pseudoscalar $Y$, the dominant cross section is the $4b$ production in the lower $Y$ mass region. For scalar $Y$ masses above the $HH$ and below the $t\bar{t}$ threshold the production of three SM-like Higgs bosons from the scalar $X$ decay $X \to H+ (Y\to HH)$ subsequently decaying into $6b$'s can dominate. Above the $t\bar{t}$ threshold the $Y$ decay into top-quark pairs takes over such that the $(b\bar{b})(t\bar{t})$ final states lead to the largest cross sections, with the exception of a few parameter points where $H+(HH)\to 6b$ still dominates (which are an artifact of different sample point densities in the two channels and should disappear with large-enough sample-size). For pseudoscalar resonant production we find a similar behavior: Above the $t\bar{t}$ threshold the $(b\bar{b})(t\bar{t})$ final state dominates. We do not have three Higgs bosons final states in the pseudoscalar case  since this channel is forbidden by the assumption of CP conservation. 

\subsection{Comparison with Previous Results}
\begin{figure}[h!]
\begin{center}
\includegraphics[width = 15cm]{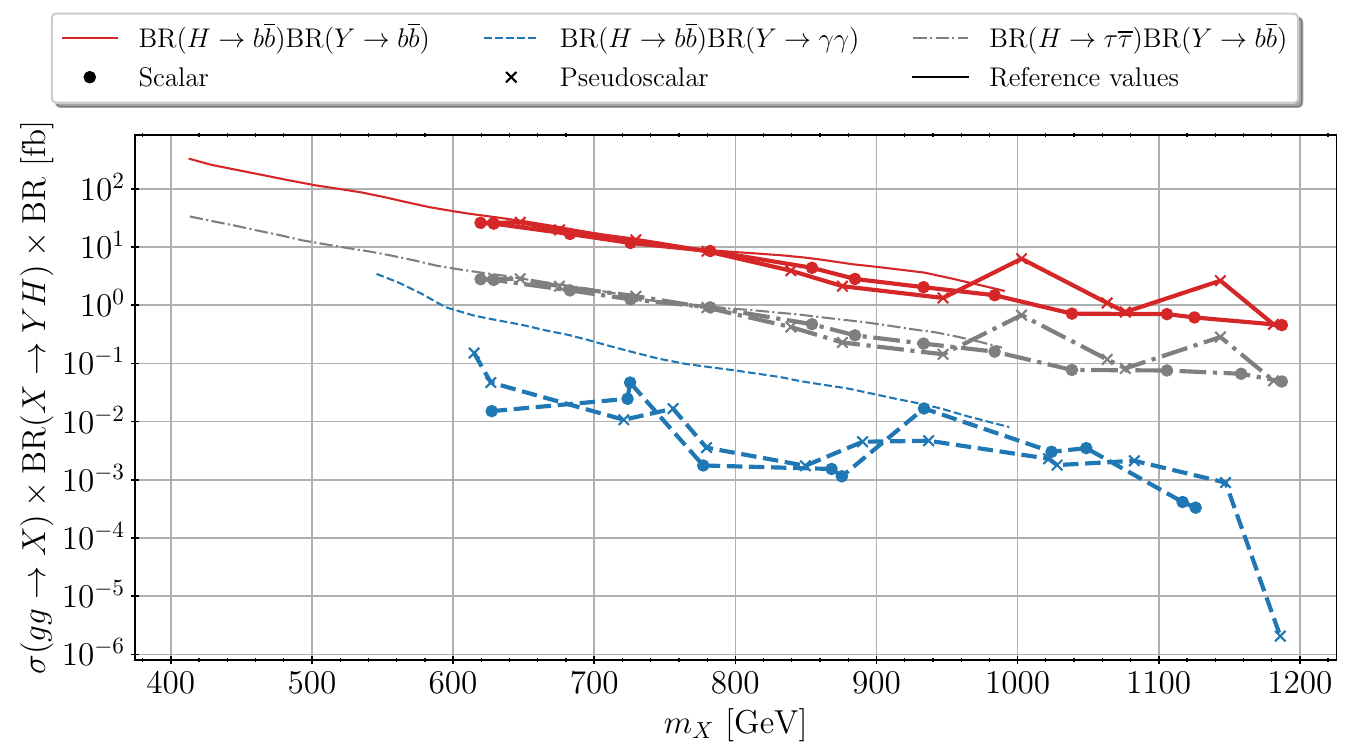}
\caption{Comparison of our results with previously produced benchmarks on maximum cross sections in the NMSSM given in Ref.~\cite{Ellwanger:2024etv} (called "reference values" in the figure).
Our values are given by the dots (crosses) and connected by lines for resonant scalar (pseudoscalar) production and compared to the reference values given in the literature (thinner line) in the $4b$ (red, full line), $(\tau\bar \tau)(b\bar b)$ (gray, dot-dashed) and $(b\bar{b})(\gamma\gamma)$ (blue, dashed) final states. The reference values stem for the former two from resonant scalar and for the latter from resonant pseudoscalar production, whereas we give our results both for resonant scalar and pseudoscalar production. Due to our chosen scan ranges our lines start above $m_X=600$~GeV.   \label{fig:comparison}}
\end{center}
\end{figure}

We compared our results with previously produced benchmarks on maximum cross sections in the NMSSM given in Refs.~\cite{Ellwanger:2022jtd} and \cite{Ellwanger:2024etv}. The comparison is shown in Fig.~\ref{fig:comparison} for the $4b$ (red), $(2\tau)(2b)$ (gray), and $(2b)(2\gamma)$ (blue) final states.
The literature results (called "reference values" in the figure) are given by the thinner lines and stem from resonant scalar production for the former two final states and from resonant pseudoscalar production for the last final state. Our results are given both for resonant scalar production (dashed lines) and for resonant pseudoscalar production (dot-dashed lines). As can be inferred from the plots, for masses above \unit[600]{GeV}, we have overall good agreement, with the exception of the $(2b)(2\gamma)$ final state, where our results are systematically lower, in particular in the lower resonant mass range $m_X$.
In general, it is difficult to get large cross sections here, as the parameter space regions e.g.~that maximize the branching ratio of the decay $A_2\to A_1 H$ and of $A_1\to \gamma\gamma$ are mutually exclusive, which is why we improved the efficiency of the scan with the modified likelihood given in \cref{eq:likelihoodsinglet} (as both branching ratios are maximized for a pure singlet/doublet $A_2/A_1$ state). Inspection of the benchmark points provided in the literature (reference values) shows that these benchmarks are characterized by very large stop mass values ranging above \unit[100]{TeV}. This kind of scenario requires a careful treatment of large logarithmic enhancements, ideally by integrating out all color-charged particles and computing the Higgs boson spectra and observables within an EFT that only includes scalars and electroweakinos. Traditional (fixed-order) calculations of the masses and mixing angles entering the Higgs boson observables are known to become increasingly unreliable for increasing stop masses \cite{Slavich:2020zjv}. For this reason, we do not allow for stop masses larger than \unit[4]{TeV} in the scan. Additionally, we check that the SM-like Higgs boson mass obtained when matching to the SM-EFT agrees with the one computed in the fixed-order calculation within \unit[3]{GeV}. 
The calculation of masses and mixing angles with an appropriate EFT, such as presented in \cite{Gabelmann:2019jvz}, is left for future works. We summarize that for a more-reliable prediction we recommend to use the benchmark scenarios provided here. 

\section{Conclusions \label{sec:summary}}
We have presented the first version of our program package \texttt{NMSSMScanner} built to perform efficient parameter scans in the complex multi-parameter space of the NMSSM. As a proof of concept we provided benchmark points that maximize the cross sections of a SM-like plus non-SM-like Higgs boson pair final state from resonant scalar or pseudoscalar production in various decay channels. On request, further benchmark points can be provided.  In an upcoming publication, the program package will be presented in full detail, including further refinements. Suggestions for the presentation of benchmark points as well as for requirements on the program package are welcome.

\section*{Acknowledgements}

We thank Mark Goodsell, Miguel Romão and Fernando Abreu de Souza for discussions.
The work of R.B. is supported in part by the Deutsche Forschungsgemeinschaft (DFG, German Research Foundation) under grant 396021762-TRR 257. MM acknowledges support by the BMBF-Project 05H24VKB.
F.E.~is supported by the DFG Emmy Noether Grant No.\ BR 6995/1-1. 
F.E.~acknowledges support by the Deutsche Forschungsgemeinschaft (DFG, German Research Foundation) under Germany's Excellence Strategy --- EXC 2121 ``Quantum Universe'' --- 390833306. F.E.'s work has been partially funded by the Deutsche Forschungsgemeinschaft (DFG, German Research Foundation) --- 491245950. 
K. E.~acknowledges financial support from the Avicenna-Studienwerk. J.P.~is grateful to support by the Studienstiftung des Deutschen Volkes. 

\begin{appendix}
\numberwithin{equation}{section}

\section{Tables for the Benchmark Points into Light Final States}
We here give the relevant information for the presented benchmark points involving an intermediate scalar or pseudoscalar resonance $X$, where the final state scalars $H$ and $Y$ decay into light final states, $H,Y \to (b\bar{b}), (\tau\tau), (\gamma\gamma)$. We remind the reader that all single and double Higgs production cross sections in this and the following section are given for a c.m.~energy of 13~TeV and that all given di-Higgs cross sections are the resonant ones.\s

The tables with the input parameters and the Higgs boson spectrum and widths as well as the relevant production cross sections and branching ratios for the benchmark points \texttt{BPs4b}, \texttt{BPs2b2gam}, \texttt{BPs2b2gam}, and \texttt{BPs2gam2b} with an intermediate scalar resonance are given in Tabs.~\ref{tab:bps4b}-\ref{tab:valuesbps2gam2b}. Those with an intermediate pseudoscalar resonance, \texttt{BPp4b} and  \texttt{BPp2b2gam}, are given in Tabs.~\ref{tab:bpp4b}-\ref{tab:valuesbpp2b2gam}.

\begin{table}[h!]
\begin{center}
\begin{tabular}{|c|c|c|c|c|c|}
\hline
$\lambda$ & $\kappa$ & $A_\lambda$ [GeV] &  $A_\kappa$ [GeV] &
$\mu_{\text{eff}}$ [GeV] & $\tan\beta$
\\ \hline
0.47 & 0.56 & 294 & -973 & 212 & 3.21
\\ \hline \hline
$m_{H^\pm}$ [GeV] & $M_1$ [GeV] & $M_2$ [GeV] & $M_3$ [GeV] & $A_t$
[GeV] & $A_b$ [GeV] \\ \hline
618.3 & 825.2 & 519.1 & 1350.5 & -3358.7 & -1372.7  \\ \hline \hline
$m_{\tilde{Q}_3}$ [GeV] & $m_{\tilde{t}_R}$ [GeV] & $m_{\tilde{b}_R}$ [GeV] & $A_\tau$
[GeV] & $m_{\tilde{L}_3}$  [GeV] & $m_{\tilde{\tau}_R}$ [GeV] \\ \hline
923.6 & 2619.9 & 3629.3 & 3847.5 & 2693.8 & 2361.8 \\ \hline
\end{tabular}

\caption{\underline{{\tt BPs4b}:} NMSSM input parameters. The soft breaking masses of the first two generations are $m_{\tilde{Q}_{1,2}}=m_{\tilde{L}_{1,2}}=3.8$~TeV, $m_{\tilde{u}_R,\tilde{d}_R}=m_{\tilde{c}_R,\tilde{s}_R}= m_{\tilde{e}_R,\tilde{\mu}_R}=3$~TeV. \label{tab:bps4b}}
\end{center}
\end{table}

\begin{table}[h!]
\begin{center}
\begin{tabular}{|c|c|c|c|c|c|}
\hline
$m_{H_1}$ [GeV] & $m_{H_2}$ [GeV] & $m_{H_3}$ [GeV] & $m_{A_1}$ [GeV] &
$m_{A_2}$ [GeV] & $m_{H^\pm}$ [GeV] \\ \hline
62.7 & 124.6 & 624.3 & 617.1 & 836.7 & 618.3
\\ \hline \hline
$\Gamma^{\text{tot}}_{H_1}$ [GeV] & $\Gamma^{\text{tot}}_{H_2}$ [GeV] &
$\Gamma^{\text{tot}}_{H_3}$ [GeV] & $\Gamma^{\text{tot}}_{A_1}$ [GeV] &
$\Gamma^{\text{tot}}_{A_2}$ [GeV] & $\Gamma^{\text{tot}}_{H^\pm}$ [GeV]
\\ \hline
2.50e-04 & 4.18e-03 & 3.97 & 4.99 & 6.30 & 4.12
\\ \hline \hline
$\sigma_{H_3}^{\text{NNLO}}$ [fb] & BR$_{H_3 \to H_1 H_2}$ & BR$_{H_1 \to b\bar{b}}$ & BR$_{H_2 \to b\bar{b}}$& $\sigma_{H_1 H_2}$ [fb] & $\sigma^s_{\text{max}}$ [fb] \\ \hline
144 & 0.336 & 0.909 & 0.617 &  48.493 & 27
  \\ \hline
\end{tabular}

\caption{\underline{{\tt BPs4b}:} The Higgs boson spectrum (upper row) with
  the total widths (middle row); the NNLO QCD $H_3$ production cross
  section, relevant branching ratios, the $H_1H_2$ and the $4b$ final
  state cross section values (lower row). The $H_1$ and $A_2$ are
  singlet-like. \label{tab:valuesbps4b}}
\end{center}
\end{table}


\begin{table}[h!]
\begin{center}
\begin{tabular}{|c|c|c|c|c|c|}
\hline
$\lambda$ & $\kappa$ & $A_\lambda$ [GeV] &  $A_\kappa$ [GeV] &
$\mu_{\text{eff}}$ [GeV] & $\tan\beta$
\\ \hline
0.55 & 0.36 & 328 & -772 & 320 & 1.64
\\ \hline \hline
$m_{H^\pm}$ [GeV] & $M_1$ [GeV] & $M_2$ [GeV] & $M_3$ [GeV] & $A_t$
[GeV] & $A_b$ [GeV] \\ \hline
605.3 & 1354.8 & 539.9 & 3769.5 & -2982.7 & -1000.9  \\ \hline \hline
$m_{\tilde{Q}_3}$ [GeV] & $m_{\tilde{t}_R}$ [GeV] & $m_{\tilde{b}_R}$ [GeV] & $A_\tau$
[GeV] & $m_{\tilde{L}_3}$  [GeV] & $m_{\tilde{\tau}_R}$ [GeV] \\ \hline
2248.3 & 2739.9 & 2176.4 & 1996.7 & 3929.5 & 1538.2 \\ \hline
\end{tabular}
\caption{\underline{{\tt BPs2b2gam}:} NMSSM input parameters. The soft breaking masses of the first two generations are $m_{\tilde{Q}_{1,2}}=m_{\tilde{L}_{1,2}}=2.7$~TeV, $m_{\tilde{u}_R,\tilde{d}_R}=m_{\tilde{c}_R,\tilde{s}_R}= m_{\tilde{e}_R,\tilde{\mu}_R}=4$~TeV. \label{tab:bps2b2gam}}
\end{center}
\end{table}

\begin{table}[h!]
\begin{center}
\begin{tabular}{|c|c|c|c|c|c|}
\hline
$m_{H_1}$ [GeV] & $m_{H_2}$ [GeV] & $m_{H_3}$ [GeV] & $m_{A_1}$ [GeV] &
$m_{A_2}$ [GeV] & $m_{H^\pm}$ [GeV] \\ \hline
125.0 & 133.1 & 611.1 & 605.4 & 701.0 & 605.3
\\ \hline \hline
$\Gamma^{\text{tot}}_{H_1}$ [GeV] & $\Gamma^{\text{tot}}_{H_2}$ [GeV] &
$\Gamma^{\text{tot}}_{H_3}$ [GeV] & $\Gamma^{\text{tot}}_{A_1}$ [GeV] &
$\Gamma^{\text{tot}}_{A_2}$ [GeV] & $\Gamma^{\text{tot}}_{H^\pm}$ [GeV]
\\ \hline
4.18$\times 10^{-3}$ & 4.66$\times 10^{-5}$ & 8.54 & 11.43 & 3.23 & 10.69
\\ \hline \hline
$\sigma_{H_3}^{\text{NNLO}}$ [fb] & BR$_{H_3 \to H_1 H_2}$ & BR$_{H_1 \to b\bar{b}}$ & BR$_{H_2 \to \gamma\gamma}$& $\sigma_{H_1 H_2}$ [fb] & $\sigma^s_{\text{max}}$ [fb] \\ \hline
617 & 0.041 & 0.613 & 7.787 $\times 10^{-3}$ & 25.025 & 0.119
  \\ \hline
\end{tabular}
\caption{\underline{{\tt BPs2b2gam}:} The Higgs boson spectrum (upper row) with the total widths (middle row); the NNLO QCD $H_3$ production cross section, relevant branching ratios, the $H_1H_2$ and the $(2b)(2\gamma)$ final state cross section values (lower row). The $H_2$ and $A_2$ are singlet-like. \label{tab:valuesbps2b2gam}}
\end{center}
\vspace*{-0.5cm}
\end{table}


\begin{table}[h!]
\begin{center}
\begin{tabular}{|c|c|c|c|c|c|}
\hline
$\lambda$ & $\kappa$ & $A_\lambda$ [GeV] &  $A_\kappa$ [GeV] &
$\mu_{\text{eff}}$ [GeV] & $\tan\beta$
\\ \hline
0.49 & 0.58 & 258 & -1054 & 229 & 2.78
\\ \hline \hline
$m_{H^\pm}$ [GeV] & $M_1$ [GeV] & $M_2$ [GeV] & $M_3$ [GeV] & $A_t$
[GeV] & $A_b$ [GeV] \\ \hline
601.0 & 951.8 & 476.9 & 1404.4 & -3456.2 & -1189.7  \\ \hline \hline
$m_{\tilde{Q}_3}$ [GeV] & $m_{\tilde{t}_R}$ [GeV] & $m_{\tilde{b}_R}$ [GeV] & $A_\tau$
[GeV] & $m_{\tilde{L}_3}$  [GeV] & $m_{\tilde{\tau}_R}$ [GeV] \\ \hline
1045.8 & 2378.6 & 3732.3 & 3879.7 & 2601.3 & 2249.4 \\ \hline
\end{tabular}
\caption{\underline{{\tt BPs2gam2b}:} NMSSM input parameters. The soft breaking masses of the first two generations are $m_{\tilde{Q}_{1,2}}=m_{\tilde{L}_{1,2}}=3.8$~TeV, $m_{\tilde{u}_R,\tilde{d}_R}=m_{\tilde{c}_R,\tilde{s}_R}= m_{\tilde{e}_R,\tilde{\mu}_R}=3.1$~TeV. \label{tab:bps2gam2b}}
\end{center}
\end{table}

\begin{table}[h!]
\begin{center}
\begin{tabular}{|c|c|c|c|c|c|}
\hline
$m_{H_1}$ [GeV] & $m_{H_2}$ [GeV] & $m_{H_3}$ [GeV] & $m_{A_1}$ [GeV] &
$m_{A_2}$ [GeV] & $m_{H^\pm}$ [GeV] \\ \hline
81.1 & 126.7 & 608.0 & 599.9 & 907.1 & 601.0
\\ \hline \hline
$\Gamma^{\text{tot}}_{H_1}$ [GeV] & $\Gamma^{\text{tot}}_{H_2}$ [GeV] &
$\Gamma^{\text{tot}}_{H_3}$ [GeV] & $\Gamma^{\text{tot}}_{A_1}$ [GeV] &
$\Gamma^{\text{tot}}_{A_2}$ [GeV] & $\Gamma^{\text{tot}}_{H^\pm}$ [GeV]
\\ \hline
6.65e-04 & 3.79e-03 & 4.33 & 5.88 & 7.59 & 4.87
\\ \hline \hline
$\sigma_{H_3}^{\text{NNLO}}$ [fb] & BR$_{H_3 \to H_1 H_2}$ & BR$_{H_1 \to b\bar{b}}$ & BR$_{H_2 \to \gamma\gamma}$& $\sigma_{H_1 H_2}$ [fb] & $\sigma^s_{\text{max}}$ [fb] \\ \hline
219 & 0.228 & 0.903 & 2.673 $\times 10^{-3}$ & 50.0 & 0.121
  \\ \hline
\end{tabular}
\caption{\underline{{\tt BPs2gam2b}:} The Higgs boson spectrum (upper row) with the total widths (middle row); the NNLO QCD $H_3$ production cross section, relevant branching ratios, the $H_1H_2$ and the $(2\gamma)(2b)$ final state cross section values (lower row). The $H_1$ and $A_2$ are singlet-like. \label{tab:valuesbps2gam2b}}
\end{center}
\end{table}

\begin{table}[h!]
\begin{center}
\begin{tabular}{|c|c|c|c|c|c|}
\hline
$\lambda$ & $\kappa$ & $A_\lambda$ [GeV] &  $A_\kappa$ [GeV] &
$\mu_{\text{eff}}$ [GeV] & $\tan\beta$
\\ \hline
0.56 & 0.57 & 106 & -65 & 346 & 1.80
\\ \hline \hline
$m_{H^\pm}$ [GeV] & $M_1$ [GeV] & $M_2$ [GeV] & $M_3$ [GeV] & $A_t$
[GeV] & $A_b$ [GeV] \\ \hline
603.2 & 894.8 & 1284.0 & 3177.1 & -3598.7 & 3886.4  \\ \hline \hline
$m_{\tilde{Q}_3}$ [GeV] & $m_{\tilde{t}_R}$ [GeV] & $m_{\tilde{b}_R}$ [GeV] & $A_\tau$
[GeV] & $m_{\tilde{L}_3}$  [GeV] & $m_{\tilde{\tau}_R}$ [GeV] \\ \hline
1582.9 & 1306.4 & 1766.9 & 3700.3 & 3060.5 & 2323.5 \\ \hline
\end{tabular}
\caption{\underline{{\tt BPp4b}:} NMSSM input parameters. The soft
  breaking masses of the first two generations are
  $m_{\tilde{Q}_{1,2}}=m_{\tilde{L}_{1,2}}=2.6$~TeV,
  $m_{\tilde{u}_R,\tilde{d}_R}=m_{\tilde{c}_R,\tilde{s}_R}=
  m_{\tilde{e}_R,\tilde{\mu}_R}=3.4$~TeV. \label{tab:bpp4b}}
\end{center}
\end{table}

\begin{table}[h!]
\begin{center}
\begin{tabular}{|c|c|c|c|c|c|}
\hline
$m_{H_1}$ [GeV] & $m_{H_2}$ [GeV] & $m_{H_3}$ [GeV] & $m_{A_1}$ [GeV] &
$m_{A_2}$ [GeV] & $m_{H^\pm}$ [GeV] \\ \hline
125.7 & 594.4 & 690.0 & 273.1 & 614.2 & 603.2
\\ \hline \hline
$\Gamma^{\text{tot}}_{H_1}$ [GeV] & $\Gamma^{\text{tot}}_{H_2}$ [GeV] &
$\Gamma^{\text{tot}}_{H_3}$ [GeV] & $\Gamma^{\text{tot}}_{A_1}$ [GeV] &
$\Gamma^{\text{tot}}_{A_2}$ [GeV] & $\Gamma^{\text{tot}}_{H^\pm}$ [GeV]
\\ \hline
4.25 $\time 10^{-3}$ & 7.05 & 3.62 & 7.26 $\time 10^{-4}$ & 9.93 & 9.71
\\ \hline \hline
$\sigma_{A_2}^{\text{NNLO}}$ [fb] &
BR$_{A_2 \to H_1 A_1}$ & BR$_{H_1 \to b\bar{b}}$ & BR$_{H_1 \to \tau\bar{\tau}}$ & BR$_{A_1 \to b\bar{b}}$ 
                & $\sigma_{H_1 A_1}$ [fb] \\ \hline
772 & 0.123 & 0.596 & 0.064 & 0.737 & 94.588   \\ \hline
\end{tabular}

\caption{\underline{{\tt BPp4b}:} The Higgs boson spectrum (upper row) with the total widths (middle row); the NNLO QCD $A_2$ production cross section, relevant branching ratios (for the $4b$ and $(\tau\bar \tau)(b\bar{b})$ final states) and the $H_1A_1$ section value (lower row). Further relevant branching ratios for the $(\gamma\gamma) (b\bar{b})$ and $(\gamma\gamma)(\tau\bar{\tau})$ final states are BR$_{H_1 \to \gamma\gamma}=2.32\times
    10^{-3}$ and BR$_{A_1 \to \tau\tau}=0.091$. The $H_3$ and $A_1$ are singlet-like.  \label{tab:valuesbpp4b}}
\end{center}
\vspace*{-0.5cm}
\end{table}


\begin{table}[h!]
\begin{center}
\begin{tabular}{|c|c|c|c|c|c|}
\hline
$\lambda$ & $\kappa$ & $A_\lambda$ [GeV] &  $A_\kappa$ [GeV] &
$\mu_{\text{eff}}$ [GeV] & $\tan\beta$
\\ \hline
0.58 & 0.58 & 243 & -38 & 330 & 1.19
\\ \hline \hline
$m_{H^\pm}$ [GeV] & $M_1$ [GeV] & $M_2$ [GeV] & $M_3$ [GeV] & $A_t$
[GeV] & $A_b$ [GeV] \\ \hline
601.6 & 419.3 & 1104.4 & 3209.0 & -3541.2 & 3769.5  \\ \hline \hline
$m_{\tilde{Q}_3}$ [GeV] & $m_{\tilde{t}_R}$ [GeV] & $m_{\tilde{b}_R}$ [GeV] & $A_\tau$
[GeV] & $m_{\tilde{L}_3}$  [GeV] & $m_{\tilde{\tau}_R}$ [GeV] \\ \hline
1473.0 & 1370.4 & 1762.2 & 3946.3 & 3173.3 & 2254.4 \\ \hline
\end{tabular}
\caption{\underline{{\tt BPp2b2gam}:} NMSSM input parameters. The soft breaking masses of the first two generations are $m_{\tilde{Q}_{1,2}}=m_{\tilde{L}_{1,2}}=2.5$~TeV, $m_{\tilde{u}_R,\tilde{d}_R}=m_{\tilde{c}_R,\tilde{s}_R}= m_{\tilde{e}_R,\tilde{\mu}_R}=3.4$~TeV. \label{tab:bpp2b2gam}}
\end{center}
\end{table}

\begin{table}[h!]
\begin{center}
\begin{tabular}{|c|c|c|c|c|c|}
\hline
$m_{H_1}$ [GeV] & $m_{H_2}$ [GeV] & $m_{H_3}$ [GeV] & $m_{A_1}$ [GeV] &
$m_{A_2}$ [GeV] & $m_{H^\pm}$ [GeV] \\ \hline
124.9 & 602.4 & 638.8 & 243.2 & 608.1 & 601.6
\\ \hline \hline
$\Gamma^{\text{tot}}_{H_1}$ [GeV] & $\Gamma^{\text{tot}}_{H_2}$ [GeV] &
$\Gamma^{\text{tot}}_{H_3}$ [GeV] & $\Gamma^{\text{tot}}_{A_1}$ [GeV] &
$\Gamma^{\text{tot}}_{A_2}$ [GeV] & $\Gamma^{\text{tot}}_{H^\pm}$ [GeV]
\\ \hline
4.13 $\times 10^{-3}$ & 14.67 & 4.62 & 1.83 $\times 10^{-4}$ & 20.73 & 19.96
\\ \hline \hline
$\sigma_{A_2}^{\text{NNLO}}$ [fb] &
BR$_{A_2 \to H_1 A_1}$ & BR$_{H_1 \to b\bar{b}}$ & BR$_{H_1 \to \tau\bar{\tau}}$ & BR$_{A_1  \to \gamma\gamma}$
                & $\sigma_{H_1 A_1}$ [fb] \\ \hline
1889 & 0.034 & 0.612 & 0.066 & 9.0 $\times 10^{-3}$ & 64.325   \\ \hline
\end{tabular}
\caption{\underline{{\tt BPp2b2gam}:} The Higgs boson spectrum (upper row) with the total widths (middle row); the NNLO QCD $A_2$ production cross section, relevant branching ratios and the $H_1A_1$ cross section value  (lower row). The $H_3$ and $A_1$ are singlet-like. 
\label{tab:valuesbpp2b2gam}}
\end{center}
\end{table}

\newpage
\section{Tables for the Benchmark Points into Heavy Final States}
Information on the benchmark points \texttt{BPs2b2t} and \texttt{BPp2b2t} with an intermediate scalar or pseudoscalar resonance $X$, respectively, where the final state SM-like $H$ decays as $H\to b\bar{b}$ and the non-SM-like Higgs boson $Y$ decays into top quarks, $Y \to t\bar{t}$, is given in Tabs.~\ref{tab:bps2b2t}-\ref{tab:valuesbpp2b2t}. The information on the benchmark point \texttt{BPs3H6b} with an intermediate scalar resonance and $H\to b\bar{b}$, $Y \to HH \to (b\bar b)(b\bar b)$ ending up in $6b$ final state is given in Tabs.~\ref{tab:bps6b} and \ref{tab:valuesbps6b}. All relevant information for the benchmark point \texttt{BPs2gamma2w} with an intermediate scalar resonance and $H\to \gamma\gamma$, $Y \to WW$ is given in Tabs.~\ref{tab:bps2gam2w} and \ref{tab:valuesbps2gam2w}.

\begin{table}[h!]
\begin{center}
\begin{tabular}{|c|c|c|c|c|c|}
\hline
$\lambda$ & $\kappa$ & $A_\lambda$ [GeV] &  $A_\kappa$ [GeV] &
$\mu_{\text{eff}}$ [GeV] & $\tan\beta$
\\ \hline
0.55 & 0.50 & 318 & -983 & 432 & 1.96
\\ \hline \hline
$m_{H^\pm}$ [GeV] & $M_1$ [GeV] & $M_2$ [GeV] & $M_3$ [GeV] & $A_t$
[GeV] & $A_b$ [GeV] \\ \hline
846.3 & 1050.2 & 1676.9 & 3188.2 & -2592.8 & 3316.8  \\ \hline \hline
$m_{\tilde{Q}_3}$ [GeV] & $m_{\tilde{t}_R}$ [GeV] & $m_{\tilde{b}_R}$ [GeV] & $A_\tau$
[GeV] & $m_{\tilde{L}_3}$  [GeV] & $m_{\tilde{\tau}_R}$ [GeV] \\ \hline
2208.9 & 3574.2 & 3712.5 & -2193.5 & 3907.3 & 3766.5 \\ \hline
\end{tabular}
\caption{\underline{{\tt BPs2b2t}:} NMSSM input parameters. The soft breaking masses of the first two generations are $m_{\tilde{Q}_{1,2}}=m_{\tilde{L}_{1,2}}=3.1$~TeV, $m_{\tilde{u}_R,\tilde{d}_R}=m_{\tilde{c}_R,\tilde{s}_R}= m_{\tilde{e}_R,\tilde{\mu}_R}=2.5$~TeV. \label{tab:bps2b2t}}
\end{center}
\end{table}

\begin{table}[h!]
\begin{center}
\begin{tabular}{|c|c|c|c|c|c|}
\hline
$m_{H_1}$ [GeV] & $m_{H_2}$ [GeV] & $m_{H_3}$ [GeV] & $m_{A_1}$ [GeV] &
$m_{A_2}$ [GeV] & $m_{H^\pm}$ [GeV] \\ \hline
124.1 & 459.9 & 852.3 & 844.2 & 1056.5 & 846.3
\\ \hline \hline
$\Gamma^{\text{tot}}_{H_1}$ [GeV] & $\Gamma^{\text{tot}}_{H_2}$ [GeV] &
$\Gamma^{\text{tot}}_{H_3}$ [GeV] & $\Gamma^{\text{tot}}_{A_1}$ [GeV] &
$\Gamma^{\text{tot}}_{A_2}$ [GeV] & $\Gamma^{\text{tot}}_{H^\pm}$ [GeV]
\\ \hline
3.98 $\times 10^{-3}$ & 0.03 & 40.44 & 11.28 & 7.89 & 11.10
\\ \hline \hline
$\sigma_{H_3}^{\text{NNLO}}$ [fb] & BR$_{H_3 \to H_1 H_2}$ & BR$_{H_1 \to b\bar{b}}$ & BR$_{H_2 \to t\bar{t}}$ & $\sigma_{H_1 H_2}$ [fb] & $\sigma^s_{\text{max}}$ [fb]  \\ \hline
71 & 0.768 & 0.614 & 0.903 & 54.375 & 30
  \\ \hline
\end{tabular}
\caption{\underline{{\tt BPs2b2t}:} The Higgs boson spectrum (upper row) with
  the total widths (middle row); the NNLO QCD $H_3$ production cross
  section, relevant branching ratios, the $H_1H_2$ cross and the $(b\bar b)(t\bar t)$ final state cross section values (lower row). The $H_2$ and $A_2$ are
  singlet-like. \label{tab:valuesbps2b2t}}
\end{center}
\end{table}

\begin{table}[h!]
\begin{center}
\begin{tabular}{|c|c|c|c|c|c|}
\hline
$\lambda$ & $\kappa$ & $A_\lambda$ [GeV] &  $A_\kappa$ [GeV] &
$\mu_{\text{eff}}$ [GeV] & $\tan\beta$
\\ \hline
0.59 & 0.60 & 62 & -151 & 344 & 2.43
\\ \hline \hline
$m_{H^\pm}$ [GeV] & $M_1$ [GeV] & $M_2$ [GeV] & $M_3$ [GeV] & $A_t$
[GeV] & $A_b$ [GeV] \\ \hline
610.5 & 1491.1 & 502.4 & 2709.1 & -3223.0 & -1672.9  \\ \hline \hline
$m_{\tilde{Q}_3}$ [GeV] & $m_{\tilde{t}_R}$ [GeV] & $m_{\tilde{b}_R}$ [GeV] & $A_\tau$
[GeV] & $m_{\tilde{L}_3}$  [GeV] & $m_{\tilde{\tau}_R}$ [GeV] \\ \hline
2940.3 & 2676.9 & 1908.6 & 2158.1 & 3759.3 & 1650.9 \\ \hline
\end{tabular}
\caption{\underline{{\tt BPp2b2t}:} NMSSM input parameters. The soft breaking masses of the first two generations are $m_{\tilde{Q}_{1,2}}=m_{\tilde{L}_{1,2}}=2.7$~TeV, $m_{\tilde{u}_R,\tilde{d}_R}=m_{\tilde{c}_R,\tilde{s}_R}= m_{\tilde{e}_R,\tilde{\mu}_R}=3.9$~TeV. \label{tab:bpp2b2t}}
\end{center}
\end{table}

\begin{table}[h!]
\begin{center}
\begin{tabular}{|c|c|c|c|c|c|}
\hline
$m_{H_1}$ [GeV] & $m_{H_2}$ [GeV] & $m_{H_3}$ [GeV] & $m_{A_1}$ [GeV] &
$m_{A_2}$ [GeV] & $m_{H^\pm}$ [GeV] \\ \hline
123.9 & 583.1 & 667.2 & 380.8 & 626.4 & 610.5
\\ \hline \hline
$\Gamma^{\text{tot}}_{H_1}$ [GeV] & $\Gamma^{\text{tot}}_{H_2}$ [GeV] &
$\Gamma^{\text{tot}}_{H_3}$ [GeV] & $\Gamma^{\text{tot}}_{A_1}$ [GeV] &
$\Gamma^{\text{tot}}_{A_2}$ [GeV] & $\Gamma^{\text{tot}}_{H^\pm}$ [GeV]
\\ \hline
3.92 $\times 10^{-3}$ & 2.47 & 2.74 & 0.16 & 5.75 & 5.77
\\ \hline \hline
$\sigma_{A_2}^{\text{NNLO}}$ [fb] &
BR$_{A_2 \to H_1 A_1}$ & BR$_{H_1 \to b\bar{b}}$ & BR$_{A_1 \to t\bar{t}}$&  $\sigma_{H_1 A_1}$ [fb] & $\sigma^p_{\text{max}}$ [fb] \\ \hline
371 & 0.165 & 0.614 & 0.977 & 61.290  & 37 \\ \hline
\end{tabular}
\caption{\underline{{\tt BPp2b2t}:} The Higgs boson spectrum (upper row) with the total widths (middle row); the NNLO QCD $A_2$ production cross section, relevant branching ratios, the $H_1A_1$ and the $(b\bar b)(t\bar t)$ cross section values (lower row). The $H_3$ and $A_1$ are singlet-like. \label{tab:valuesbpp2b2t}}
\end{center}
\vspace*{-0.5cm}
\end{table}

\begin{table}[h!]
\begin{center}
\begin{tabular}{|c|c|c|c|c|c|}
\hline
$\lambda$ & $\kappa$ & $A_\lambda$ [GeV] &  $A_\kappa$ [GeV] &
$\mu_{\text{eff}}$ [GeV] & $\tan\beta$
\\ \hline
0.57 & 0.60 & 93 & -1146 & 332 & 2.11
\\ \hline \hline
$m_{H^\pm}$ [GeV] & $M_1$ [GeV] & $M_2$ [GeV] & $M_3$ [GeV] & $A_t$
[GeV] & $A_b$ [GeV] \\ \hline
601.0 & 416.4 & 530.9 & 3619.6 & -3830.3 & -1757.8  \\ \hline \hline
$m_{\tilde{Q}_3}$ [GeV] & $m_{\tilde{t}_R}$ [GeV] & $m_{\tilde{b}_R}$ [GeV] & $A_\tau$
[GeV] & $m_{\tilde{L}_3}$  [GeV] & $m_{\tilde{\tau}_R}$ [GeV] \\ \hline
3220.0 & 3868.4 & 2756.0 & 1707.5 & 3168.4 & 1937.6 \\ \hline
\end{tabular}
\caption{\underline{{\tt BPs6b}:} NMSSM input parameters. The soft breaking masses of the first two generations are $m_{\tilde{Q}_{1,2}}=m_{\tilde{L}_{1,2}}=2.9$~TeV, $m_{\tilde{u}_R,\tilde{d}_R}=m_{\tilde{c}_R,\tilde{s}_R}= m_{\tilde{e}_R,\tilde{\mu}_R}=3.6$~TeV. \label{tab:bps6b}}
\end{center}
\end{table}

\begin{table}[h!]
\begin{center}
\begin{tabular}{|c|c|c|c|c|c|}
\hline
$m_{H_1}$ [GeV] & $m_{H_2}$ [GeV] & $m_{H_3}$ [GeV] & $m_{A_1}$ [GeV] &
$m_{A_2}$ [GeV] & $m_{H^\pm}$ [GeV] \\ \hline
124.0 & 291.5 & 610.4 & 599.0 & 1059.3 & 601.0
\\ \hline \hline
$\Gamma^{\text{tot}}_{H_1}$ [GeV] & $\Gamma^{\text{tot}}_{H_2}$ [GeV] &
$\Gamma^{\text{tot}}_{H_3}$ [GeV] & $\Gamma^{\text{tot}}_{A_1}$ [GeV] &
$\Gamma^{\text{tot}}_{A_2}$ [GeV] & $\Gamma^{\text{tot}}_{H^\pm}$ [GeV]
\\ \hline
3.82e-03 & 0.15 & 5.54 & 7.32 & 12.51 & 6.97
\\ \hline \hline
$\sigma_{H_3}^{\text{NNLO}}$ [fb] & BR$_{H_3 \to H_1 H_2}$ & BR$_{H_1 \to b\bar{b}}$ & BR$_{H_2 \to b\bar{b}}$& BR$_{H_2 \to H_1 H_1}$& $\sigma_{H_1 H_2}$ [fb]  \\ \hline
374 & 0.110 & 0.608 & 0.006 & 0.435 & 41.214
  \\ \hline
\end{tabular}
\caption{\underline{{\tt BPs6b}:} The Higgs spectrum (upper row) with
  the total widths (middle row); the NNLO QCD $H_3$ production cross
  section, relevant branching ratios and the resulting $H_1H_2$ cross-section (lower row). The $H_1$ and $A_2$ are
  singlet-like. \label{tab:valuesbps6b}}
\end{center}
\vspace*{-0.5cm}
\end{table}

\begin{table}[h!]
\begin{center}
\begin{tabular}{|c|c|c|c|c|c|}
\hline
$\lambda$ & $\kappa$ & $A_\lambda$ [GeV] &  $A_\kappa$ [GeV] &
$\mu_{\text{eff}}$ [GeV] & $\tan\beta$
\\ \hline
0.63
& 0.51
 & 234
 & -931
 & 326
 & 1.86
\\ \hline \hline
$m_{H^\pm}$ [GeV] & $M_1$ [GeV] & $M_2$ [GeV] & $M_3$ [GeV] & $A_t$
[GeV] & $A_b$ [GeV] \\ \hline
 601.8
 & 1367.0
 & 553.8
 & 3625.9
 & -3011.3
 &  -1105.9
 \\ \hline \hline
$m_{\tilde{Q}_3}$ [GeV] & $m_{\tilde{t}_R}$ [GeV] & $m_{\tilde{b}_R}$ [GeV] & $A_\tau$
[GeV] & $m_{\tilde{L}_3}$  [GeV] & $m_{\tilde{\tau}_R}$ [GeV] \\ \hline
2202.1
 &  2844.5
 & 2240.8
 & 2224.3
 & 3812.2
 & 1446.5
 \\ \hline
\end{tabular}
\caption{\underline{{\tt BPs2gamma2w}:} NMSSM input parameters. The soft breaking masses of the first two generations are $m_{\tilde{Q}_{1,2}}=m_{\tilde{L}_{1,2}}=2.55$~TeV, $m_{\tilde{u}_R,\tilde{d}_R}=m_{\tilde{c}_R,\tilde{s}_R}= m_{\tilde{e}_R,\tilde{\mu}_R}=4.0$~TeV. \label{tab:bps2gam2w}}
\end{center}
\end{table}

\begin{table}[h!]
\begin{center}
\begin{tabular}{|c|c|c|c|c|c|}
\hline
$m_{H_1}$ [GeV] & $m_{H_2}$ [GeV] & $m_{H_3}$ [GeV] & $m_{A_1}$ [GeV] &
$m_{A_2}$ [GeV] & $m_{H^\pm}$ [GeV] \\ \hline
125.6
 & 178.6  & 610.1 & 602.5 & 844.1 & 601.8
\\ \hline \hline
$\Gamma^{\text{tot}}_{H_1}$ [GeV] & $\Gamma^{\text{tot}}_{H_2}$ [GeV] &
$\Gamma^{\text{tot}}_{H_3}$ [GeV] & $\Gamma^{\text{tot}}_{A_1}$ [GeV] &
$\Gamma^{\text{tot}}_{A_2}$ [GeV] & $\Gamma^{\text{tot}}_{H^\pm}$ [GeV]
\\ \hline
 3.45 $\times 10^{-3}$ &  5.72 $\times 10^{-2}$ & 6.98 & 9.14
  & 7.83
 & 8.78
\\ \hline \hline
$\sigma_{H_3}^{\text{NNLO}}$ [fb] & BR$_{H_3 \to H_1 H_2}$ & BR$_{H_1 \to  \gamma\gamma}$ & BR$_{H_2 \to WW}$ & $\sigma_{H_1 H_2}$ [fb] & $\sigma^s_{\text{max}}$ [fb] \\ \hline
 479 & 8.69 $\times 10^{-2}$  &   2.68 $\times 10^{-3}$
 & 0.93 & 41.625 & 0.104
  \\ \hline
\end{tabular}
\caption{\underline{{\tt BPs2gamma2w}:} The Higgs boson spectrum (upper row) with
  the total widths (middle row); the NNLO QCD $H_3$ production cross
  section, relevant branching ratios, the $H_1H_2$ and the $(2\gamma)(2W)$ final state cross section values (lower row). The $H_2$ and $A_2$ are singlet-like. \label{tab:valuesbps2gam2w}}
\end{center}
\vspace*{-0.5cm}
\end{table}
\end{appendix}

\clearpage

\bibliography{nmssmscanner.bib}

\end{document}